\documentclass[english]{article}
\usepackage[T1]{fontenc}
\usepackage[latin9]{inputenc}
\usepackage{geometry}
\usepackage{color}
\usepackage{amsmath}
\usepackage{graphicx}
\usepackage{esint}
\usepackage{ulem}
\usepackage{subfigure}
\usepackage{cite}
\usepackage{epsfig}
\usepackage{epstopdf}
\makeatletter

\newcommand{\lyxmathsym}[1]{\ifmmode\begingroup\def\b@ld{bold}
  \text{\ifx\math@version\b@ld\bfseries\fi#1}\endgroup\else#1\fi}

\providecommand{\tabularnewline}{\\}

\usepackage{cite,times}

\makeatother

\usepackage{babel}
\begin{document}

\title{Lower- and higher-order nonclassical properties of photon added and
subtracted displaced Fock states}

\author{Priya Malpani$^{\dagger}$, Nasir Alam$^{\ddagger}$, Kishore Thapliyal$^{\ddagger}$,
Anirban Pathak$^{\ddagger}$, V. Narayanan$^{\dagger}$,\\
 Subhashish Banerjee$^{\dagger}$ \\
 $^{\dagger}$Indian Institute of Technology Jodhpur, Jodhpur 342011,
India\\
 $^{\ddagger}$Jaypee Institute of Information Technology, A-10, Sector-62,
Noida, UP-201307, India}
\maketitle
\begin{abstract}
Nonclassical properties of photon added and subtracted displaced Fock
states have been studied using various witnesses of lower- and higher-order
nonclassicality. Compact analytic expressions are obtained for the
nonclassicality witnesses. Using those expressions, it is established
that these states and the states that can be obtained as {their
limiting cases} (except coherent states) are highly nonclassical as
they show the existence of lower- and higher-order antibunching and
sub-Poissonian photon statistics, in addition to the nonclassical
features revealed through the Mandel $Q_{M}$ parameter, zeros of
$Q$ function, Klyshko's criterion, and Agarwal-Tara criterion. Further,
some comparison between the nonclassicality of photon added and subtracted
displaced Fock states have been performed using witnesses of nonclassicality.
This has established that between the two types of non-Gaussianity
inducing operations (i.e., photon addition and subtraction) used here,
photon addition influences the nonclassical properties more strongly.
Further, optical designs for the generation of photon added and subtracted
displaced Fock states from squeezed vacuum state have also been proposed. 
\end{abstract}

\section{Introduction \label{sec:Intro}}

With the advent of quantum state engineering \cite{vogel1993quantum,sperling2014quantum,miranowicz2004dissipation,marchiolli2004engineering}
and quantum information processing (\cite{pathak2013elements} and
references therein), the study of nonclassical properties of engineered
quantum states have become a very important field. This is so because
the presence of nonclassical features in a quantum state can only
provide quantum supremacy. In the recent past, various techniques
for quantum state engineering have been developed \cite{vogel1993quantum,sperling2014quantum,miranowicz2004dissipation,agarwal1991nonclassical,lee2010quantum,marchiolli2004engineering}.
If we restrict ourselves to optics, these techniques are primarily
based on the clever use of beam splitters and detectors, measurements
with post selection, etc. Such techniques are useful in creating holes
in the photon number distribution \cite{escher2004controlled} and
in generating finite dimensional quantum states \cite{miranowicz2004dissipation},
both of which are nonclassical \cite{pathak2017classical}. The above
said techniques are also useful in realizing non-Gaussianity inducing
operations, like photon addition and subtraction \cite{zavatta2004quantum,podoshvedov2014extraction}.
Keeping this in mind, in what follows, we aim to study the nonclassical
properties of a set of engineered quantum states (both photon added
and subtracted) which can be produced by using the above mentioned
techniques. The work is further motivated by the fact that recently
several applications of nonclassical states have been reported. Specifically,
squeezed states have played an important role in {{}
the studies related to phase diffusion} \cite{banerjee2007phase,banerjee2007phaseQND},
the detection of gravitational waves in LIGO experiments \cite{abbasi2013thermal,abbott2016gw151226,abbott2016observation},
teleportation of coherent states \cite{furusawa1998unconditional},
and continuous variable quantum cryptography \cite{hillery2000quantum}.
The rising demand for a single photon source can be fulfilled by an
antibunched light source \cite{yuan2002electrically}. {The
study of quantum correlations is important both from the perspective
of pure and mixed states} \cite{chakrabarty2010study,dhar2013controllable,banerjee2010dynamics,banerjee2010entanglement}.
Entangled states are found to be useful in both secure \cite{ekert1991quantum}
and insecure \cite{bennett1992communication,bennett1993teleporting}
quantum communication schemes. Stronger quantum correlation present
in the steerable states are used to ensure the security against all
the side-channel attacks on devices used in one-side (i.e., either
preparation or detector side) for quantum cryptography \cite{branciard2012one}.
Quantum supremacy in computation is established due to quantum algorithms
for unsorted database search \cite{grover1997quantum}, factorization
and discrete logarithm problems \cite{shor1999polynomial}, and machine
learning \cite{biamonte2017quantum} using essentially nonclassical
states.

\begin{figure}[h]
\centering{}

\includegraphics[scale=0.7,angle=-90]{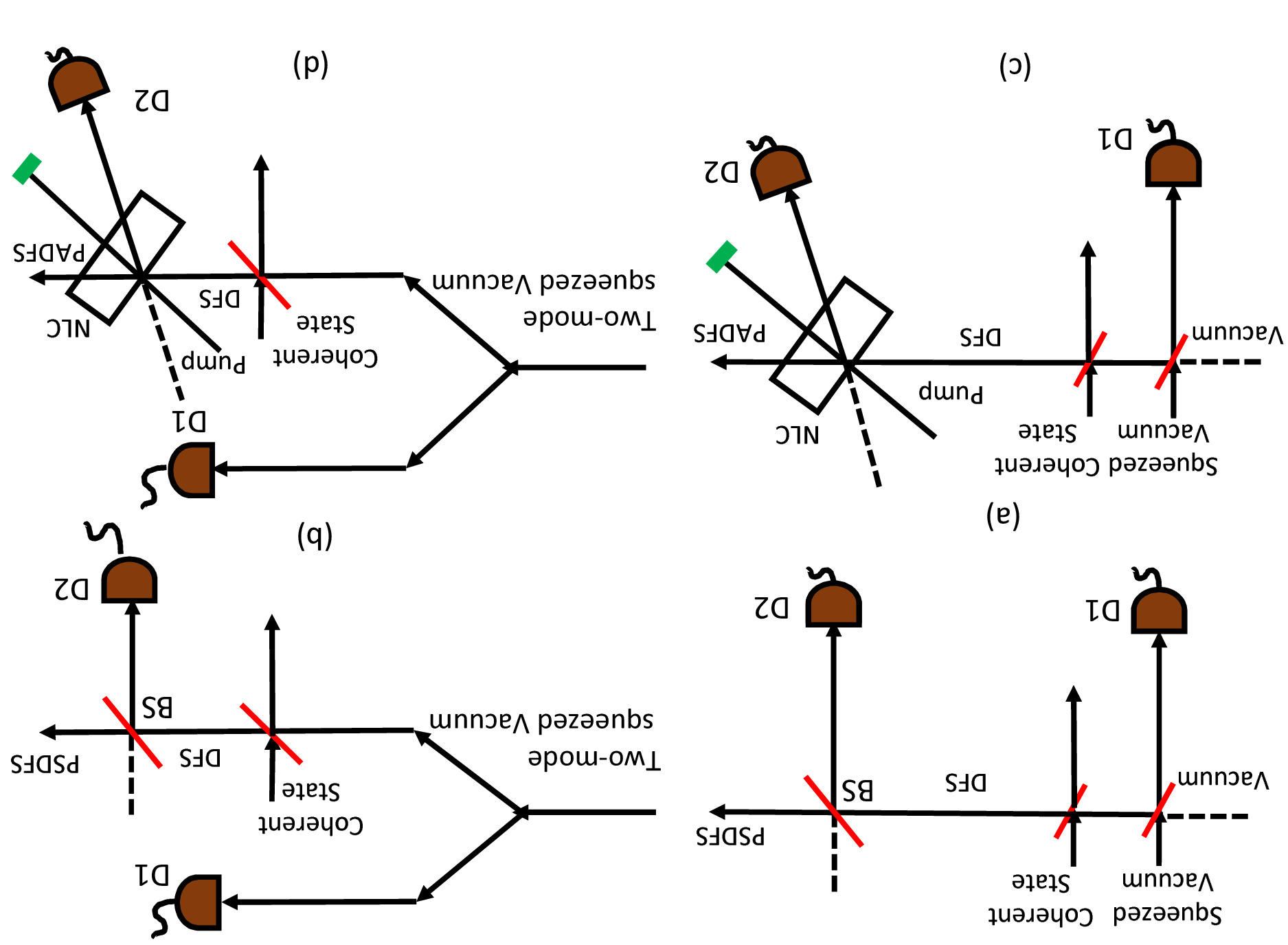} \caption{\label{DFS}(Color online) A schematic diagram for the generation
of PSDFS (in (a) and (b)) and PADFS (in (c) and (d)). In (a) and (c)
((b) and (d)), single-mode (two-mode) squeezed vacuum is used for
generation of the desired state. Here, NLC corresponds to nonlinear
crystal, BS is a beamsplitter, and D1 and D2 are photon number resolving
detectors.}
\end{figure}

It is well-known that a state having a negative $P$-function is referred
to as a nonclassical state. Such a state cannot be expressed as a
mixture of coherent states and does not possess a classical analogue.
In contrast to these states, coherent states are classical, but neither
their finite dimensional versions \cite{miranowicz2004dissipation,alam2017higher}
nor their generalized versions are classical\cite{banerjee2007phase},
\cite{satyanarayana1985generalized,thapliyal2016tomograms,thapliyal2015quasiprobability}.
Here, we would like to focus on photon added and subtracted versions
of a particular type of generalized coherent state, which is also
referred to as the displaced Fock state (DFS). To be precise, state
of the form $|\psi\rangle=D(\alpha)|n\rangle$, where $D(\alpha)$
is the displacement operator with Fock state $|n\rangle$, is referred
to as generalized coherent state (cf. Ref. \cite{satyanarayana1985generalized}),
as this state is reduced to a coherent state in the limit $n=0$.
However, from the structure of the state it seems more appropriate
to call this state as the DFS, and {this seems to
be the nomenclature usually adopted in the literature} \cite{keil2011classical,wunsche1991displaced,blavzek1994high,moya1995generation},
this state has been referred to as the DFS. In some other works, it
is referred to as displaced number state \cite{ziesel2013experimental,de2006nonlinear,dodonov2005decoherence},
but all these names are equivalent; and in what follows, we will refer
to it as DFS. This is an extremely interesting quantum state for various
reasons. Specifically, its relevance in various areas of quantum optics
is known. For example, in the context of cavity QED, it constitutes
the eigenstates of the Jaynes-Cummings systems with coherently driven
atoms \cite{alsing1992dynamic}. Naturally, various lower-order nonclassical
properties and a set of other quantum features of DFS have already
been studied. Specifically, quasiprobability distributions of DFS
were studied in \cite{wunsche1991displaced}, phase fluctuations of
DFS was investigated in \cite{zheng1992fluctuation}, decoherence
of superposition of DFS was discussed in \cite{dodonov2005decoherence},
$Q$ function, Wigner function and probability distribution of DFS
were studied in \cite{de1990properties}, Pancharatnam phase of DFS
has been studied in \cite{mendas1993pancharatnam}. Further, in the
context of non-optical DFS, various possibilities of generating DFS
from Fock states by using a general driven time-dependent oscillator
has been discussed in \cite{lo1991generating}; and in the trapped-ion
system, quantum interference effects have been studied for the superposition
of DFS \cite{marchiolli2004engineering}. Thus, DFS seems to be a
well studied quantum state, but it may be noted that {a little
effort} has yet been made to study higher-order nonclassical properties
of DFS. It is a relevant observation as in the recent past, it has
been observed that higher-order nonclassicality has various applications
\cite{hillery1999quantum,banerjee2017quantum,sharma2017quantumauction,thapliyal2017protocols},
and it can be used to detect the existence of weak nonclassical characters
\cite{thapliyal2014higher,thapliyal2014nonclassical,verma2010generalized,thapliyal2017comparison,alam2016nonclassical,alam2015approximate,thapliyal2017nonclassicality,prakash2006higher,prakash2010detection}.
Further, various higher-order nonclassical features have been experimentally
detected \cite{allevi2012measuring,allevi2012high,avenhaus2010accessing,perina2017higher}.
However, we do not want to restrict to DFS, rather we wish to focus
on lower- and higher-order nonclassical properties of a set of even
more general states, namely photon added DFS (PADFS) and photon subtracted
DFS (PSDFS). The general nature of these states can be visualized
easily, as in the special case that no photon is added (subtracted)
PADFS (PSDFS) would reduce to DFS. Further, for $n=0$, PADFS would
reduce to photon added coherent state which has been widely studied
(\cite{agarwal1991nonclassical,verma2008higher,thapliyal2017comparison}
and references therein) and experimentally realized \cite{zavatta2004quantum,zavatta2005single}.
Here it is worth noting that DFS has also been generated experimentally
by superposing a Fock state with a coherent state on a beam splitter
\cite{lvovsky2002synthesis}. Further, an alternative method for the
generation of DFS has been proposed by Oliveira et al. \cite{de2005alternative}.
From the fact that photon added coherent state and DFS have already
been generated experimentally, and the fact that the photon added
states can be prepared via conditional measurement on a beam splitter,
it appears that PADFS and PSDFS can also be built in the lab. In fact,
inspired by these experiments, we have proposed a schematic diagram
for generating the PADFS and PSDFS in Fig. \ref{DFS} using single-mode
and two-mode squeezed vacuum states. Specifically, using three (two)
highly transmitting beamsplitters, a conditional measurement of single
photons at both detectors D1 and D2 in Fig. \ref{DFS} (a) (Fig. \ref{DFS}
(b)) would result in single photon subtracted DFS as output from a
single-mode (two-mode) squeezed vacuum state. Similarly, to generate
PADFS conditional subtraction of photon is replaced by photon addition,
using a nonlinear crystal and heralding one output mode for a successful
measurement of a single photon to ensure generation of single photon
added DFS. This fact, their general nature, and the fact that nonclassical
properties of PADFS and PSDFS have not yet been {accorded
sufficient attention, has} motivated us to perform this study.

Motivated by the above facts, in what follows, we investigate the
possibilities of observing lower- and higher-order sub-Poissonian
photon statistics, antibunching and squeezing in PADFS and PSDFS.
We have studied nonclassical properties of these states through a
set of other witnesses of nonclassicality, e.g., zeros of $Q$ function,
Mandel $Q_{M}$ parameter, Klyshko's criterion, and Agarwal-Tara parameter.
These witnesses of nonclassicality successfully establish that both
PADFS and PSDFS (along with most of the states to which these two
states reduce at different limits) are highly nonclassical. Thus,
{making use of the analytical expressions
of moments of creation and annihilation operators, discussed below,
facilitates an analytical understanding for most of the nonclassical
witnesses.}

The rest of the paper is organized as follows. In Section \ref{sec:Quantum-states-of},
quantum states of our interest (i.e., PADFS and PSDFS) have been introduced
in detail. In Section \ref{sec:Nonclassicality-witnesses}, we describe
various witnesses and measures of nonclassicality and report analytic
expressions of most of them for the quantum states of our interest.
Further, the existence of various lower- and higher-order nonclassical
features in PADFS and PSDFS are shown through a set of plots. Finally,
we conclude in Section \ref{sec:Conclusions}.

\section{Quantum states of our interest \label{sec:Quantum-states-of}}

We have already mentioned that we are interested in PADFS and PSDFS.
Before mathematically defining these states, we may note that in
the Fock state basis, a DFS may be described as follows

\begin{equation}
|\phi(n,\alpha)\rangle=D(\alpha)|n\rangle=\frac{1}{\sqrt{n!}}\sum_{p=0}^{n}{n \choose p}(-\alpha^{\star})^{(n-p)}\exp\left[-\frac{\mid\alpha\mid^{2}}{2}\right]\sum_{m=0}^{\infty}\frac{\alpha^{m}}{m!}\sqrt{(m+p)!}|m+p\rangle.\label{eq:GCS}
\end{equation}
Using this state, we can define a $u$ photon added DFS (i.e., a PADFS)
as

\begin{equation}
|\psi_{+}(u,n,\alpha)\rangle=N_{+}\hat{a}^{\dagger u}|\phi(n,\alpha)\rangle=\frac{1}{\sqrt{n!}}\sum_{p=0}^{n}{n \choose p}(-\alpha^{\star})^{(n-p)}\exp\left(-\frac{\mid\alpha\mid^{2}}{2}\right)\sum_{m=0}^{\infty}\frac{\alpha^{m}}{m!}\sqrt{(m+p+u)!}|m+p+u\rangle.\label{eq:PADFS}
\end{equation}
Similarly, a $v$ photon subtracted DFS (i.e., a PSDFS) can be expressed
as 

\begin{equation}
|\psi_{-}(v,n,\alpha)\rangle=N_{-}\hat{a}^{v}|\phi(n,\alpha)\rangle=\frac{1}{\sqrt{n!}}\sum_{p=0}^{n}{n \choose p}(-\alpha^{\star})^{(n-p)}\exp\left(-\frac{\mid\alpha\mid^{2}}{2}\right)\sum_{m=0}^{\infty}\frac{\alpha^{m}}{m!}\dfrac{(m+p)!}{\sqrt{(m+p-v)!}}|m+p-v\rangle,\label{eq:PSDFS}
\end{equation}
where $m$ and $p$ are the real integers. Here, 
\begin{equation}
N_{+}=\left[\frac{1}{n!}\sum_{p,p'=0}^{n}{n \choose p}{n \choose p'}(-\alpha^{\star})^{(n-p)}(-\alpha)^{(n-p')}\exp\left[-\mid\alpha\mid^{2}\right]\sum_{m=0}^{\infty}\frac{\alpha^{m}(\alpha^{\star})^{m+p-p'}(m+p+u)!}{m!(m+p-p')!}\right]^{-\frac{1}{2}},\label{eq:norP}
\end{equation}
and 
\begin{equation}
N_{-}=\left[\frac{1}{n!}\sum_{p,p'=0}^{n}{n \choose p}{n \choose p'}\left(\alpha^{\star}\right)^{(n-p)}\left(-\alpha\right){}^{(n-p')}\exp\left[-\mid\alpha\mid^{2}\right]\sum_{m=0}^{\infty}\frac{\alpha^{m}\left(\alpha^{\star}\right){}^{m+p-p'}(m+p)!}{m!(m+p-p')!(m+p-v)!}\right]^{-\frac{1}{2}}\label{eq:norS}
\end{equation}
are the normalization constants, and subscripts $+$ and $-$ correspond
to photon addition and subtraction. Thus, $|\psi_{+}(u,n,\alpha)\rangle$
and $|\psi_{-}(v,n,\alpha)\rangle$ represent $u$ photon added DFS
and $v$ photon subtracted DFS, respectively, for the DFS which has
been produced by displacing the Fock state $|n\rangle$ by a displacement
operator $D(\alpha)$ characterized by the complex parameter $\alpha.$
Clearly, the addition and the subtraction of photons on the DFS can
be mathematically viewed as application of the creation and annihilation
operators from the left on the Eq. (\ref{eq:GCS}). Here it may be
noted that different well-known states can be obtained as special
cases of these two states. For example, using the notation introduced
above to define PADFS and PSDFS, we can describe a coherent state
$|\alpha\rangle$ as $|\alpha\rangle=|\psi_{+}(0,\alpha,0)\rangle=|\psi_{-}(0,\alpha,0)\rangle$,
naturally, coherent state can be viewed as a special case of both
PADFS and PSDFS. Similarly, we can describe a single photon added
coherent state as $|\psi\rangle_{{\rm PACS}}=|\psi_{+}(1,\alpha,0)\rangle$,
a Fock state as $|n\rangle=|\psi_{+}(0,0,n)\rangle$ and a DFS as
$|\psi\rangle_{{\rm DFS}}=|\psi_{+}(0,\alpha,n)\rangle=|\psi_{-}(0,\alpha,n)\rangle.$

In what follows, we will see that various experimentally measurable
nonclassicality witnesses can be expressed as the moments of annihilation
and creation operators \cite{allevi2012measuring,allevi2012high,avenhaus2010accessing,perina2017higher,miranowicz2010testing}.
To utilize those witnesses to identify the signatures of nonclassicality,
we will compute an analytic expression for the most general
moment, $\langle\hat{a}^{\dagger q}\hat{a}^{r}\rangle$, with $q$
and $r$ being non-negative integers. This is the most general moment
in the sense that any other moment can be obtained as a special case
of it. For example, if we need $\langle a^{2}\rangle,$ we would just
require to consider $q=2$ and $r=0$. Thus, an analytic expression
for $\langle\hat{a}^{\dagger q}\hat{a}^{r}\rangle$ would essentially
help us to obtain analytic expression for any moment-based witness
of nonclassicality. Further, the analytic expressions of moment obtained
for PADFS and PASDFS would also help us to obtain nonclassical features
in the set of states obtained in the limiting cases, like Fock state,
DFS, photon added coherent state. Keeping this in mind, we have computed
$\langle\psi_{+}(u,n,\alpha)|\hat{a}^{\dagger q}\hat{a}^{r}|\psi_{+}(u,n,\alpha)\rangle$
and $\langle\psi_{-}(v,n,\alpha)|\hat{a}^{\dagger q}\hat{a}^{r}|\psi_{-}(v,n,\alpha)\rangle$,
and provide the final analytic expressions of these moments without
going into the mathematical details to maintain the flow of the paper.
The obtained expressions for the above mentioned moments for PADFS
and PSDFS are

\begin{equation}
\begin{array}{lcl}
\langle\hat{a}^{\dagger q}\hat{a}^{r}\rangle_{{\rm PADFS}}=\langle\psi_{+}(u,n,\alpha)|\hat{a}^{\dagger q}\hat{a}^{r}|\psi_{+}(u,n,\alpha)\rangle & = & \frac{N_{+}^{2}}{n!}\sum\limits _{p,p'=0}^{n}{n \choose p}{n \choose p'}(-\alpha^{\star})^{(n-p)}(-\alpha)^{(n-p')}\exp\left[-\mid\alpha\mid^{2}\right]\\
 & \times & \sum\limits _{m=0}^{\infty}\frac{\alpha^{m}(\alpha^{\star})^{m+p-p'-r+q}(m+p+u)!(m+p+u-r+q)!}{m!(m+p-p'-r+q)!(m+p+u-r)!},
\end{array}\label{eq:PA-expepectation}
\end{equation}
and

\begin{equation}
\begin{array}{lcl}
\langle\hat{a}^{\dagger q}\hat{a}^{r}\rangle_{{\rm PSDFS}}=\langle\psi_{-}(v,n,\alpha)|\hat{a}^{\dagger q}\hat{a}^{r}|\psi_{-}(v,n,\alpha)\rangle & = & \frac{N_{-}^{2}}{n!}\sum\limits _{p,p'=0}^{n}{n \choose p}{n \choose p'}(-\alpha^{\star})^{(n-p)}(-\alpha)^{(n-p')}\exp\left[-\mid\alpha\mid^{2}\right]\\
 & \times & \sum\limits _{m=0}^{\infty}\frac{\alpha^{m}(\alpha^{*})^{m+p-p'-r+q}(m+p)!(m+p-r+q)!}{m!(m+p-p'-r+q)!(m+p-v-r)!},
\end{array}\label{eq:PS-expectation}
\end{equation}
respectively. The values of normalization constants for PADFS and
PSDFS are already given in Eqs. (\ref{eq:norP}) and (\ref{eq:norS}),
respectively. In the following section, we shall investigate the possibilities
of observing various types lower- and higher-order nonclassical features
in PADFS and PSDFS by using Eqs. (\ref{eq:PA-expepectation}) and
(\ref{eq:PS-expectation}).

\section{Nonclassical features of PADFS and PSDFS \label{sec:Nonclassicality-witnesses}}

The moments of number operators for PADFS and PSDFS states obtained
in the previous section enable us to study nonclassical properties
of these states using a set of moments-based criteria of nonclassicality
\cite{miranowicz2010testing}\cite{naikoo2018probing}. In the recent
past, an infinite set of these moments-based criteria is shown to
be equivalent to the $P$ function-based criterion, i.e., it becomes
both necessary and sufficient \cite{richter2002nonclassicality,shchukin2005nonclassical}.
However, in this section, we will use a subset of this infinite set
as witnesses of nonclassicality to investigate various nonclassical
properties of the PADFS and PSDFS. Specifically, nonclassicality will
be witnessed through Mandel $Q_{M}$ parameter, criteria of lower-
and higher-order antibunching, Agarwal-Tara criterion, Klyshko's criterion,
criteria of higher-order sub-Poissonian photon statistics, etc. We
will further discuss a quasiprobability distribution, namely $Q$
function, which is also a witness of nonclassicality. Specifically,
zeros of $Q$ function are signatures of nonclassicality.

\subsection{Mandel $Q_{M}$ Parameter}

The Mandel $Q_{M}$ parameter \cite{mandel1979sub} illustrates the
nonclassicality through photon number distribution in a quantum state.
The Mandel $Q_{M}$ parameter is defined as 
\begin{equation}
Q_{M}=\frac{\langle(\hat{a}^{\dagger}\hat{a})^{2}\rangle-\langle\hat{a}^{\dagger}\hat{a}\rangle^{2}-\langle\hat{a}^{\dagger}\hat{a}\rangle}{\langle\hat{a}^{\dagger}\hat{a}\rangle}.\label{eq:MandelQ}
\end{equation}
The negative values of $Q_{M}$ parameter essentially indicate the
negativity for $P$ function and so it gives a witness for nonclassicality
which can be calculated using Eqs. (\ref{eq:PA-expepectation}) and
(\ref{eq:PS-expectation}). For the Poissonian statistics it becomes
0, while for the sub-Poissonian (super-Poissonian) photon statistics
it has negative (positive) values.

\begin{figure}
\centering{}

\subfigure[]{\includegraphics[scale=0.6]{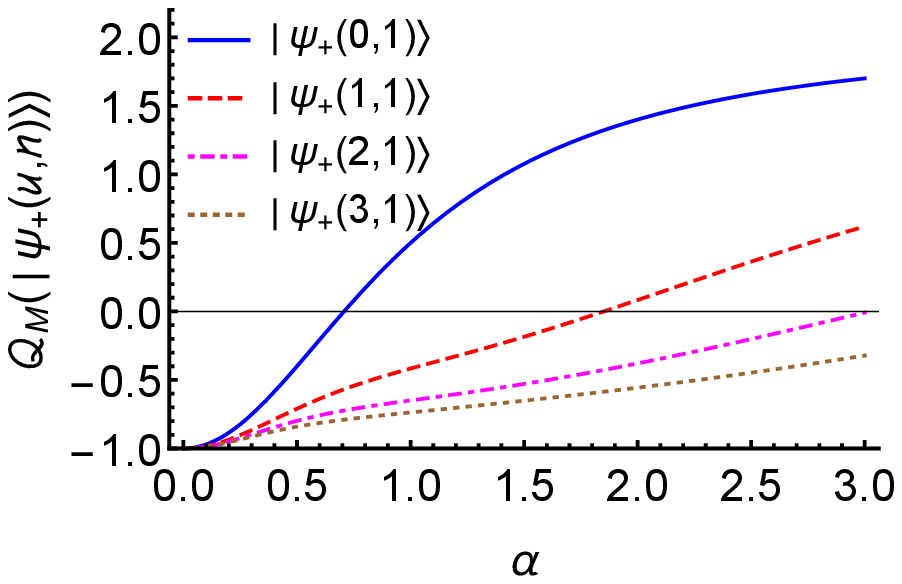}}
\quad{}\quad{}\subfigure[]{ \includegraphics[scale=0.6]{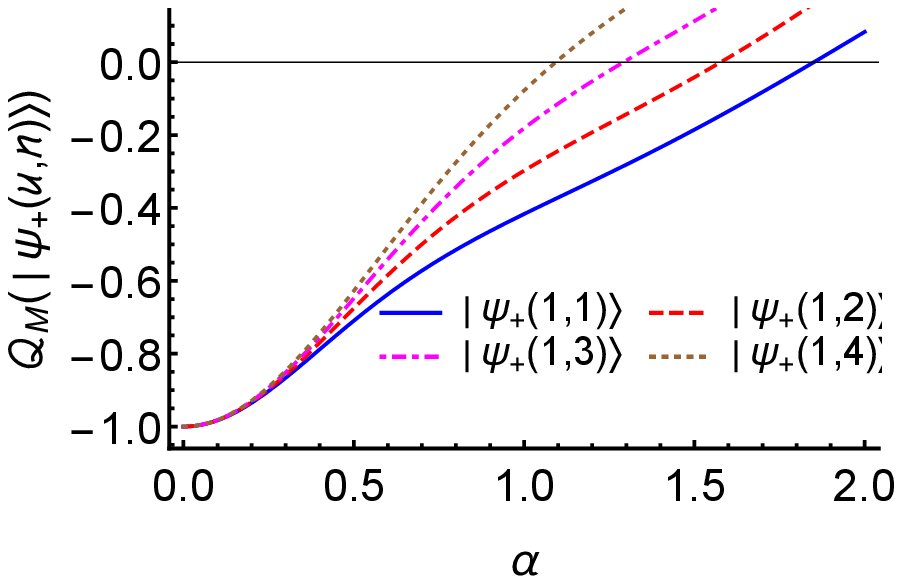}}\\
 \subfigure[]{\includegraphics[scale=0.6]{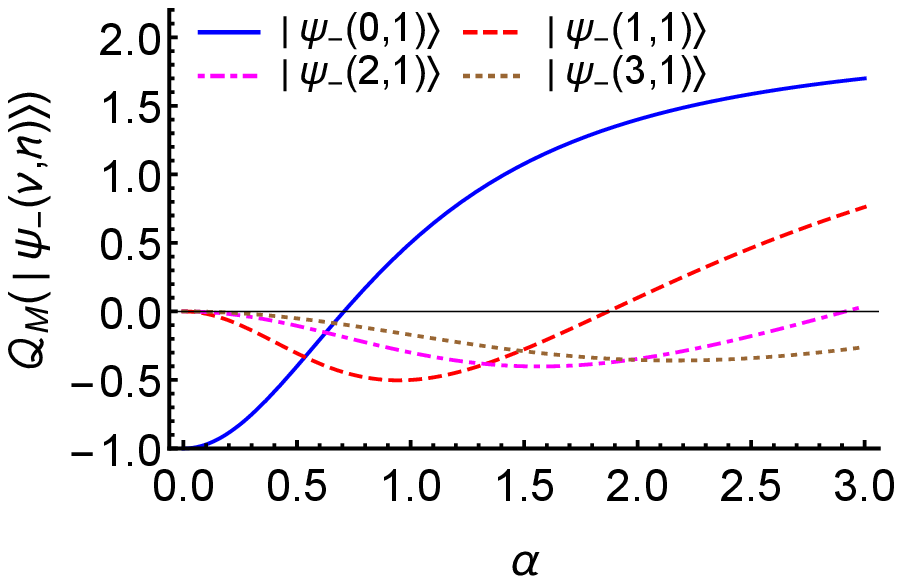}}\quad{}\quad{}\subfigure[]{
\includegraphics[scale=0.6]{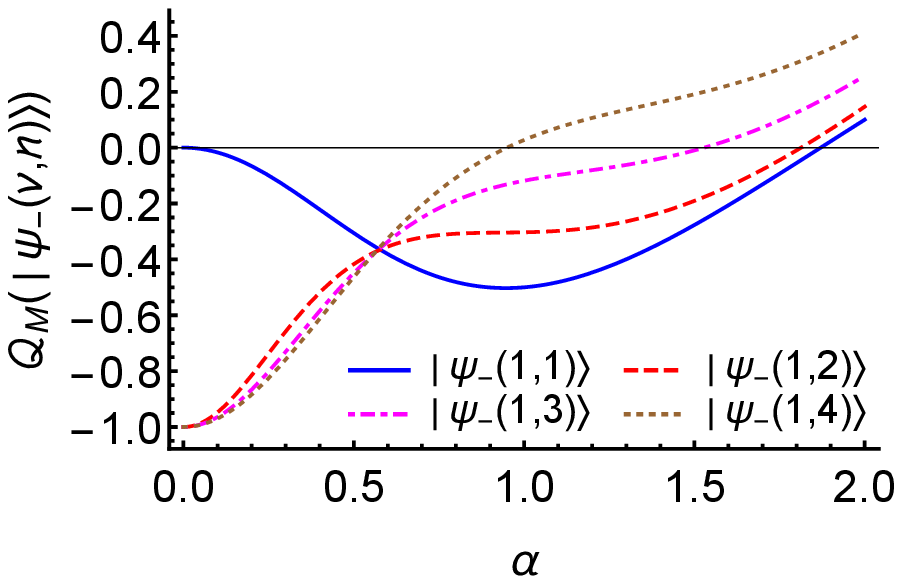}} \caption{\label{Mandel-Q-parameter}(Color online) Variation of Mandel $Q_{M}$
parameter for PADFS (in (a) and (b)) and PSDFS (in (c) and (d)) is
shown with displacement parameter $\alpha$. In (a) and (c), the value
of number of photons added/subtracted (i.e., $u$ or $v$) is changed
for the same initial Fock state $|1\rangle$. Different initial Fock
states $|n\rangle$ are chosen to be displaced in (b) and (d) for
the single photon addition/subtraction. }
\end{figure}

In Fig. \ref{Mandel-Q-parameter}, the dependence of $Q_{M}$ on the
state parameter $\alpha$ and non-Gaussianity inducing parameters
is shown. Specifically, variation of $Q_{M}$ parameter for PADFS
and PSDFS is shown with state parameter $\alpha$, where the effect
of the number of photons added/subtracted and the initial Fock state
is also established. For $\alpha=0$, the PADFS with an arbitrary
number of photon addition has $Q_{M}$ parameter -1, which can be
attributed to the fact that final state, which reduces to the Fock
state ($|1\rangle$ chosen to be displaced in this case) is the most
nonclassical state (cf. Fig. \ref{Mandel-Q-parameter} (a)). With
increase in the number of photons added to the DFS, the depth of nonclassicality
witness $Q_{M}$ increases. However, the witness of nonclassicality
tends towards the positive side for higher values of the displacement
parameter. In contrast to the photon addition, with the subtraction
of photons from the DFS the $Q_{M}$ parameter becomes almost zero
for the smaller values of displacement parameter $\alpha$ in DFS
as shown in Fig. \ref{Mandel-Q-parameter} (c). This behavior can
be attributed to the fact that photon subtraction from $D\left(\alpha\right)|1\rangle$
for small values of $\alpha$ will most likely yield vacuum state.
Also, with the increase in the displacement parameter the witness
of nonclassicality becomes more negative as with a higher average
number of photons in DFS photon subtraction becomes more effective.
However, for the larger values of displacement parameter the nonclassicality
disappears analogous to the PADFS.

As Fock states are known to be nonclassical, and photon addition and
subtraction are established as nonclassicality inducing operations,
it would be worth comparing the effect of these two independent factors
responsible for the observed nonclassical features in the present
case. To perform this study, we have shown the variation of the single
photon added (subtracted) DFS with different initial Fock states in
Fig. \ref{Mandel-Q-parameter} (b) (Fig. \ref{Mandel-Q-parameter}
(d)). Specifically, the nonclassicality present in PADFS decays faster
for the higher values of the Fock states with increasing displacement
parameter (cf. Fig. \ref{Mandel-Q-parameter} (b)). However, such
nature was not present in PSDFS shown in Fig. \ref{Mandel-Q-parameter}
(d). Note that variation of $Q_{M}$ parameter with $\alpha$ starts
from 0 (-1) iff $u\leq n$ $\left(u>n\right)$. For instance, if $u=n=1$,
i.e., corresponding to state $\hat{a}D\left(\alpha\right)|1\rangle$,
nonclassicality witness is zero for $\alpha=0$ as it corresponds
to vacuum state. Therefore, the present study reveals that photon
addition is a stronger factor for the nonclassicality present in the
state when compared to the initial Fock state chosen to be displaced.
Whereas photon subtraction is a preferred choice for large values
of displacement parameter in contrast to the higher values of Fock
states to displace with small $\alpha$. Among photon addition and
subtraction, addition is a preferred choice for the smaller values
of displacement parameter, while the choice between addition and subtraction
becomes immaterial for large $\alpha$.

\subsection{Lower- and higher-order antibunching}

\begin{figure}
\centering{}

\subfigure[]{\includegraphics[scale=0.6]{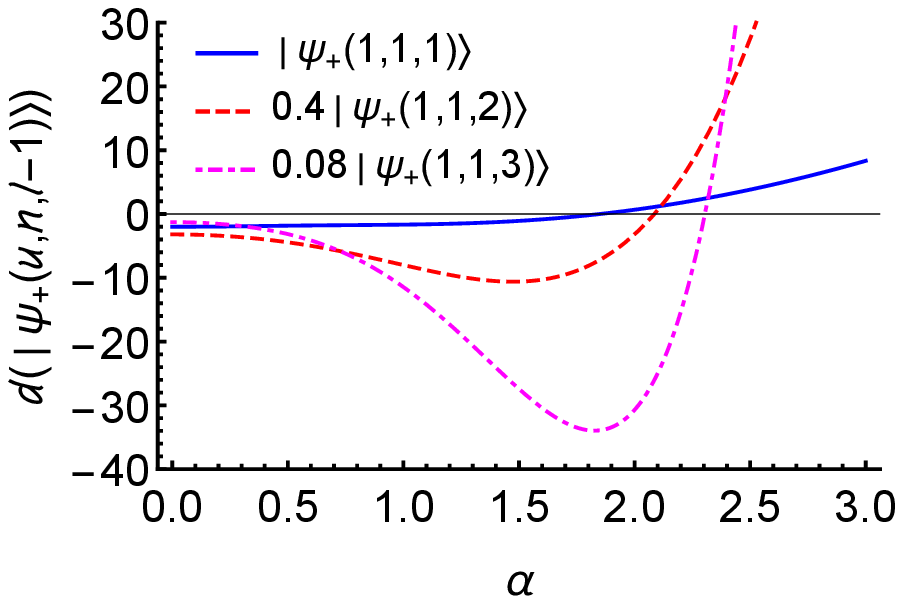}} \quad{}\subfigure[]{
\includegraphics[scale=0.6]{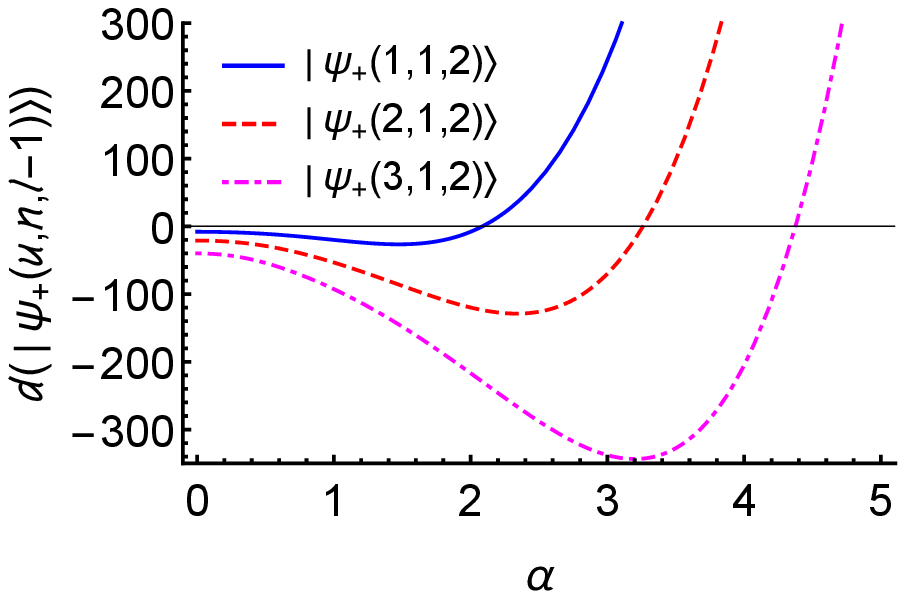}}\quad{}\subfigure[]{
\includegraphics[scale=0.6]{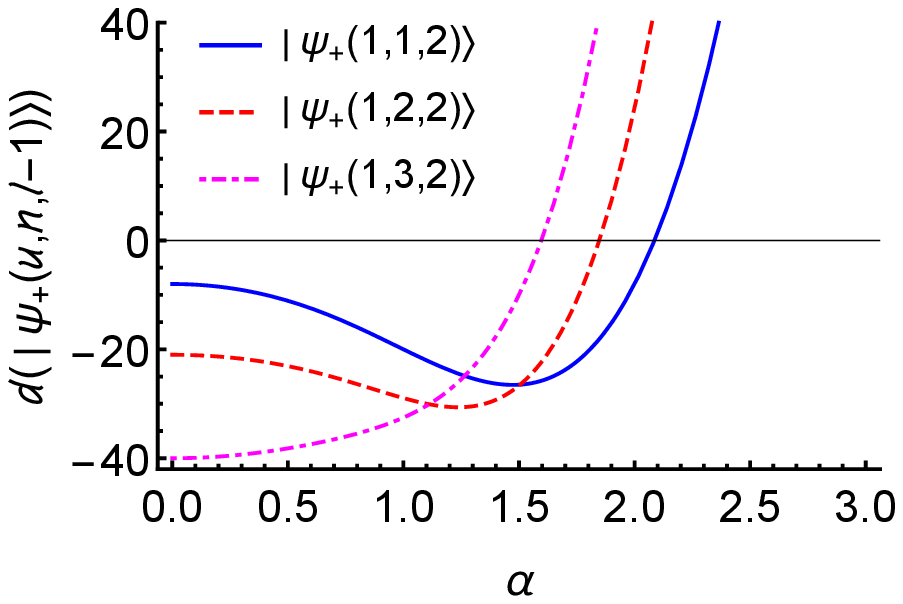}}\\
 \subfigure[]{\includegraphics[scale=0.6]{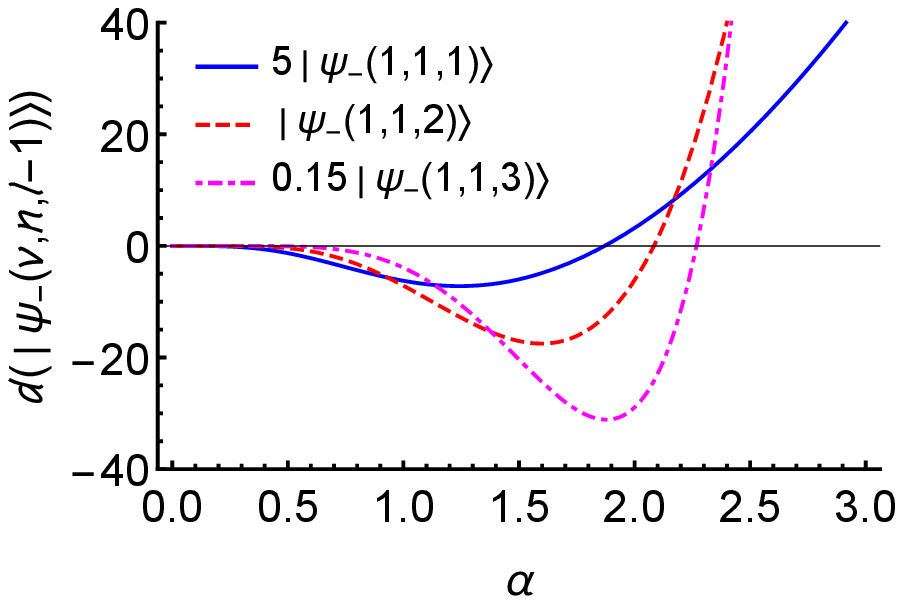}} \quad{}\subfigure[]{
\includegraphics[scale=0.6]{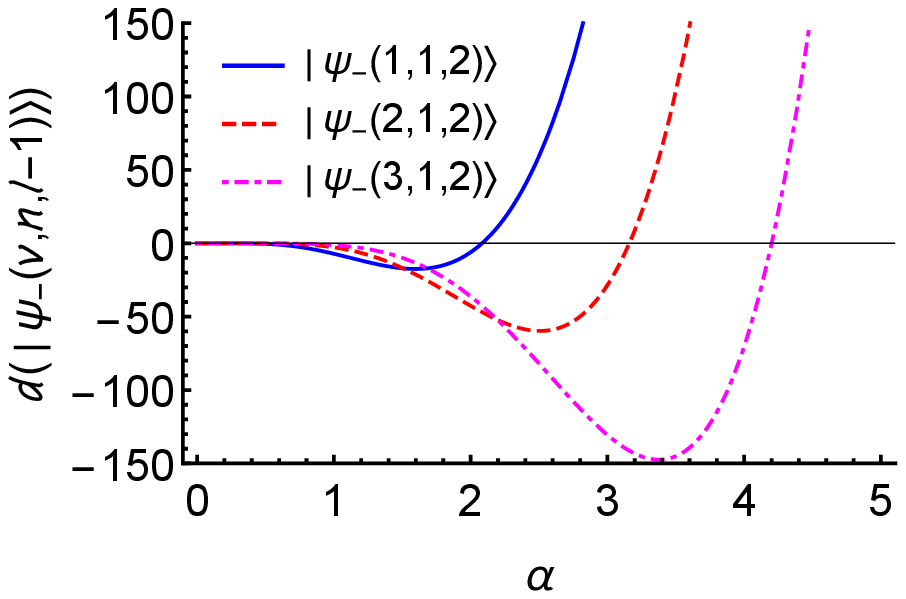}}\quad{}\subfigure[]{
\includegraphics[scale=0.6]{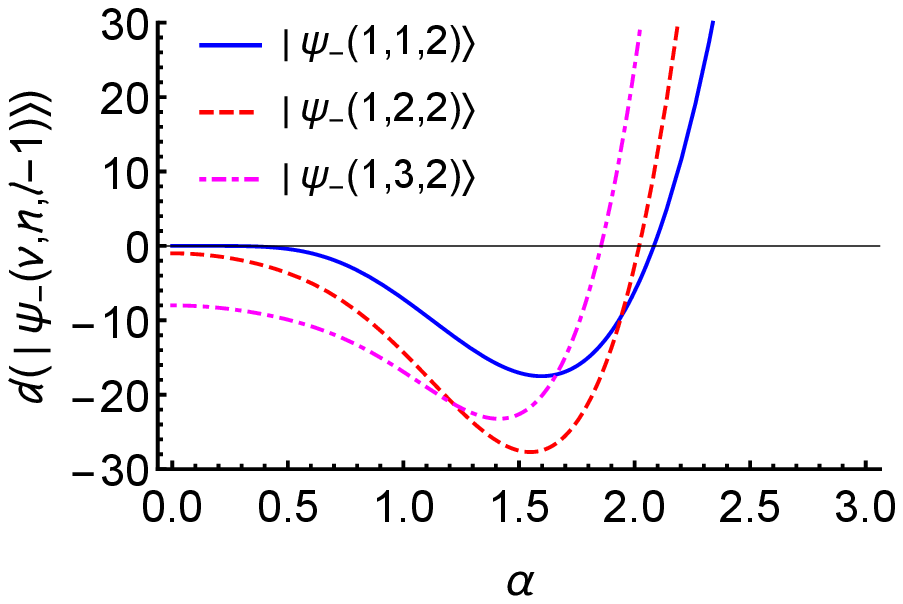}}
\caption{\label{HOA}(Color online) The presence of higher-order antibunching
is shown as a function of $\alpha$ for PADFS (in (a)-(c)) and PADFS
((d)-(f)). Specifically, (a) and (d) illustrate comparison between
lower- and higher-order antibunching. {It should be
noted that some of the curves are multiplied by a scaling factor in
order to present them in one figure}. Figures (b) and (e) show the
effect of photon addition/subtraction, and (c) and (f) establish the
effect of Fock state chosen to displace in PADFS and PSDFS, respectively.}
\end{figure}

In this section, we study lower- and higher-order antibunching. To
do so, we use the following criterion of $(l-1)$th order antibuncing
(\cite{pathak2006control} and references therein) 
\begin{equation}
d(l-1)=\langle\hat{a}^{\dagger l}\hat{a}^{l}\rangle-\langle\hat{a}^{\dagger}\hat{a}\rangle^{l}<0.\label{eq:HOA}
\end{equation}
This nonclassicality feature characterizes suitability of the quantum
state to be used as single photon source as the negative values of
$d(l-1)$ parameter shows that the probability of photons coming bunched
is less compared to that of coming independently. The signature of
lower-order antibunching can be obtained as a special case of Eq.
(\ref{eq:HOA}) for $l=2$, and that for $l\geq3$, the negative values
of $d(l-1)$ correspond to higher-order antibunching of $(l-1)$th
order. The nonclassicality reflected by the lower-order antibunching
criterion obtained here is the same as Mandel $Q_{M}$ parameter illustrated
in Fig. \ref{Mandel-Q-parameter}. Therefore, we will rather discuss
here the possibility of observing higher-order antibunching in the
quantum states of our interest using Eqs. (\ref{eq:PA-expepectation})
and (\ref{eq:PS-expectation}) in Eq. (\ref{eq:HOA}). Specifically,
the depth of nonclassicality witness can be observed to increase with
order for both PADFS and PSDFS as depicted in Fig. \ref{HOA} (a)
and (d). This fact is consistent with the earlier observations (\cite{thapliyal2014higher,thapliyal2014nonclassical,thapliyal2017comparison,thapliyal2017nonclassicality,alam2017lower}
and references therein) that higher-order nonclassicality criteria
are useful in detecting weaker nonclassicality. On top of that the
higher-order antibunching can be observed for larger values of displacement
parameter $\alpha$, when lower-order antibunching is not present.
The presence of higher-order nonclassicality in the absence of its
lower-order counterpart establishes the relevance of the present study.

The depth of nonclassicality parameter was observed to decrease with
an increase in the number of photons subtracted from DFS for small
values of $\alpha$ in Fig. \ref{Mandel-Q-parameter} (c). A similar
nature is observed in Fig. \ref{HOA} (e), which shows that for the
higher values of displacement parameter, the depth of higher-order
nonclassicality witness increases with the number of photon subtraction.
Therefore, not only the depth of nonclassicality but the range of
displacement parameter for the presence of higher-order antibunching
also increases with photon addition/subtraction (cf. Fig. \ref{HOA}
(b) and (e)). With the increase in the value of Fock state parameter
$n$, the depth of higher-order nonclassicality witness increases
(decreases) for smaller (larger) values of displacement parameter
in both PADFS and PSDFS as shown in Fig. \ref{HOA} (c) and (f), respectively.
Thus, we have observed that the range of $\alpha$ with the presence
of nonclassicality increases (decreases) with photon addition/subtraction
(Fock state) in DFS.

\subsection{Higher-order sub-Poissonian photon statistics}

The lower-order counterparts of antibunching and sub-Poissonian photon
statistics are closely associated as the presence of latter ensures
the possibility of observing former (see \cite{thapliyal2014higher,thapliyal2017comparison}
for a detailed discussion). However, these two nonclassical features
were shown to be independent phenomena in the past (\cite{thapliyal2014higher,thapliyal2017comparison}
and references therein). Higher-order counterpart of sub-Poissonian
photon statistics can be introduced as

\begin{equation}
\begin{array}{lcccc}
\mathcal{D}_{h}(l-1) & = & \sum\limits _{e=0}^{l}\sum\limits _{f=1}^{e}S_{2}(e,\,f)\,^{l}C_{e}\,\left(-1\right)^{e}d(f-1)\langle N\rangle^{l-e} & < & 0,\end{array}\label{eq:hosps22}
\end{equation}
where $S_{2}(e,\,f)$ is the Stirling number of second kind, {and
$\,^{l}C_{e}$ is the usual binomial coefficient}. This allows
us to study higher-order sub-Poissonian photon statistics using Eqs.
(\ref{eq:PA-expepectation}) and (\ref{eq:PS-expectation}) in Eq.
(\ref{eq:hosps22}).

\begin{figure}
\centering{}

\subfigure[]{\includegraphics[scale=0.6]{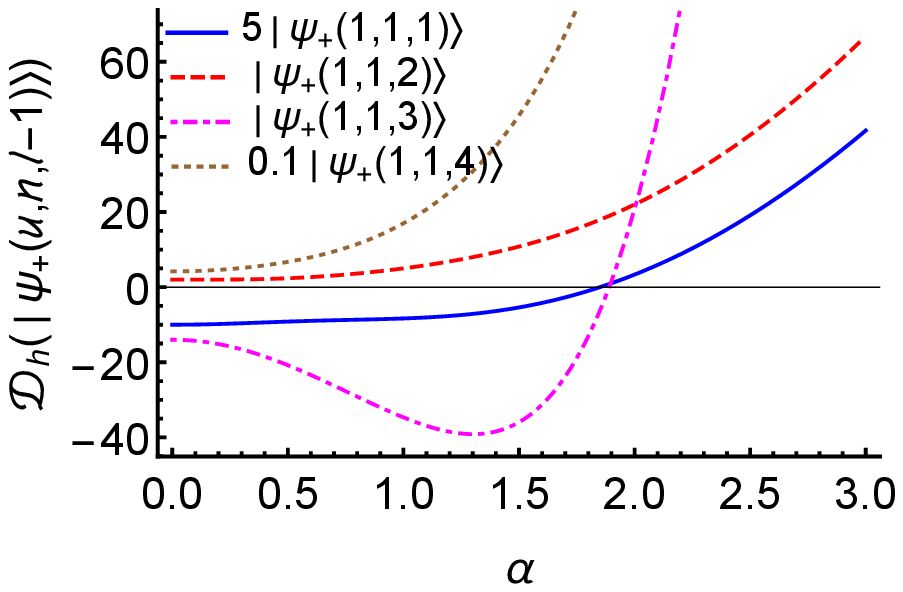}} \quad{}\subfigure[]{
\includegraphics[scale=0.6]{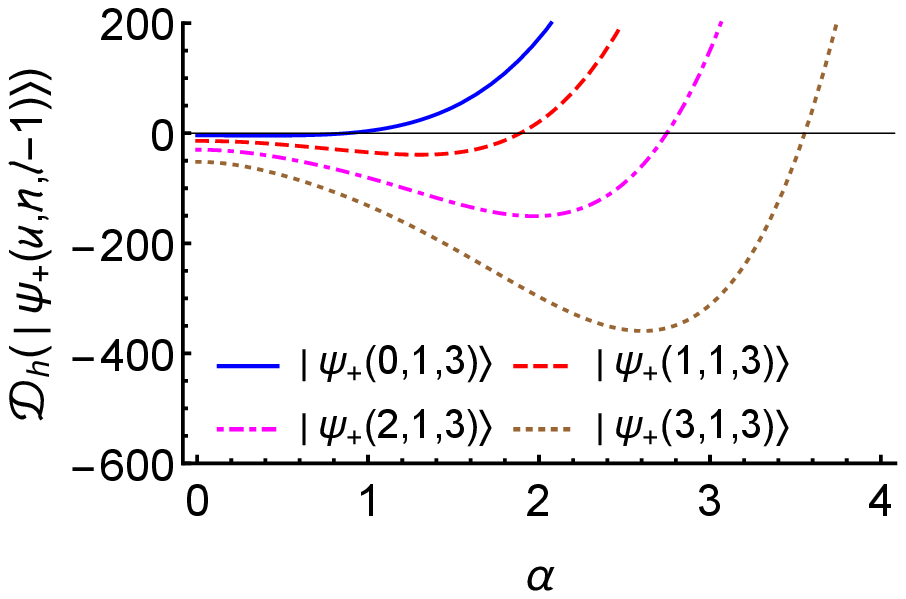}}\quad{}\subfigure[]{
\includegraphics[scale=0.6]{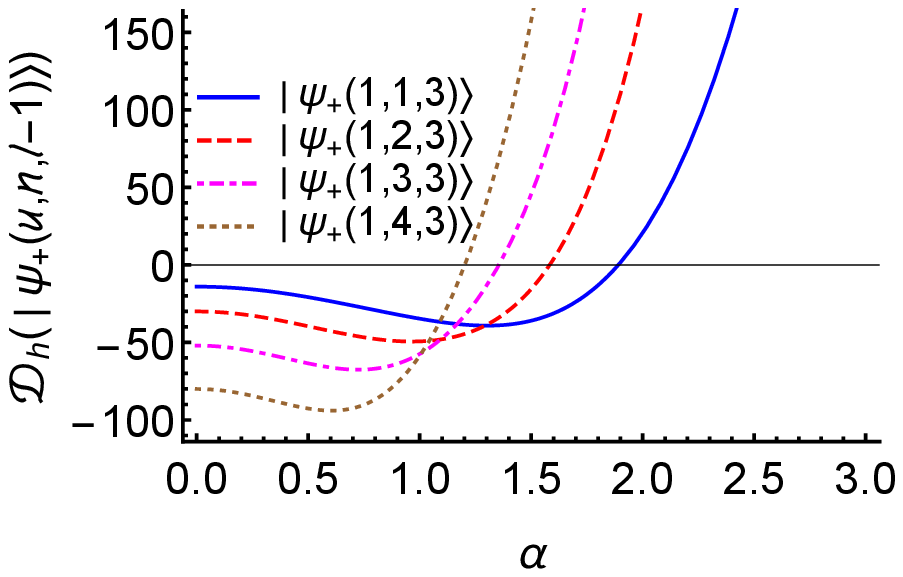}}\\
\subfigure[]{\includegraphics[scale=0.6]{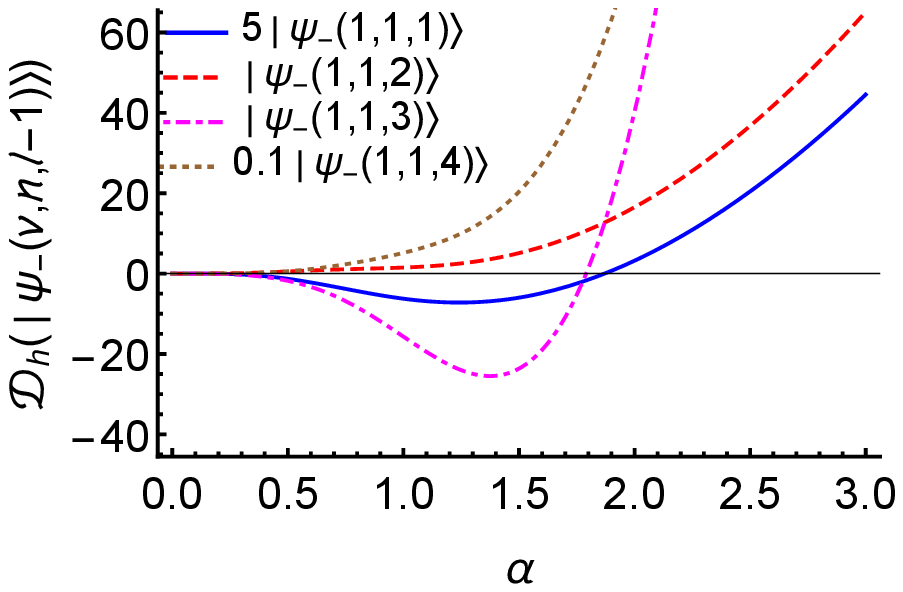}}\quad{}\subfigure[]{
\includegraphics[scale=0.6]{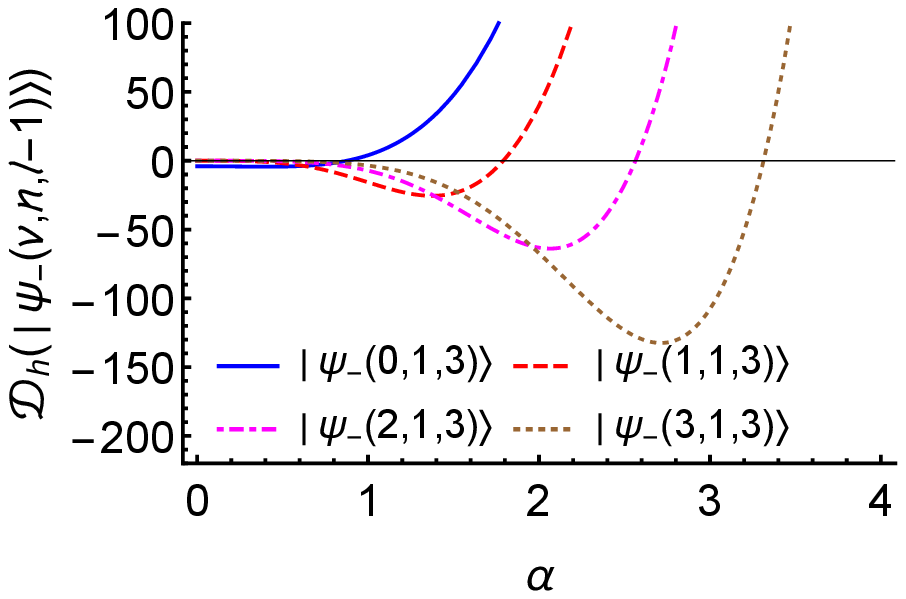}} \quad{}\subfigure[]{
\includegraphics[scale=0.6]{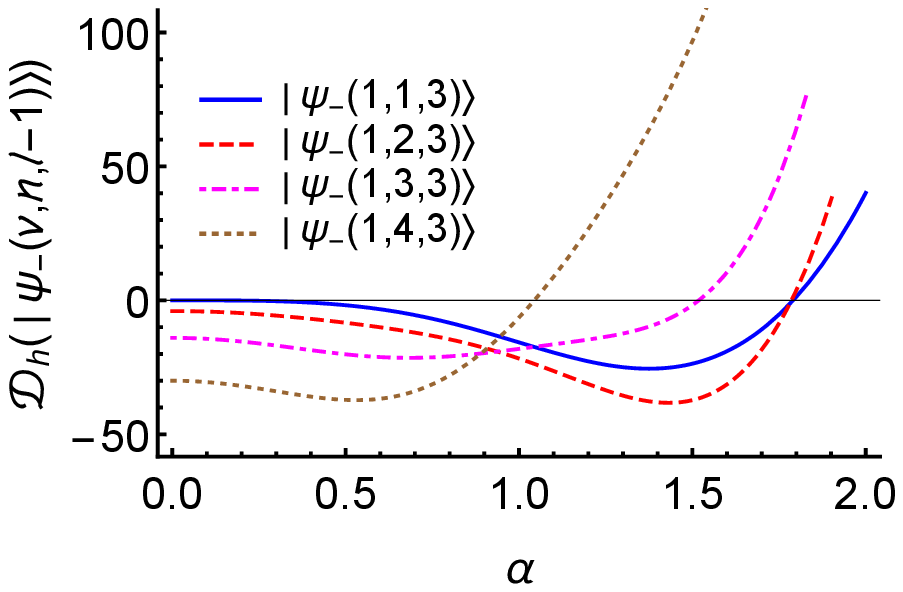}}\\

\caption{\label{HOSPS}(Color online) Dependence of higher-order sub-Poissonian
photon statistics on $\alpha$ for PADFS (in (a)-(c)) and PSDFS ((d)-(f)) is illustrated here. Specifically,
(a) and (d) show the increase in depth of nonclassicality witness
with order, (b) and (e) depict the effect of photon addition and subtraction, respectively,
and (c) and (f) establish the effect of choice of Fock state to be
displaced in PADFS and PSDFS, respectively. }
\end{figure}

The presence of higher-order sub-Poissonian photon statistics (as
can be seen in Fig. \ref{HOSPS} (a) and (d) for PADFS and PSDFS,
respectively) is dependent on the order of nonclassicality unlike
higher-order antibunching, which is observed for all orders. Specifically,
the nonclassical feature was observed only for odd orders, which is
consistent with some of our earlier observations \cite{thapliyal2017comparison},
where nonclassicality in those cases could be induced due to squeezing.
Along the same line, we expect to observe the nonclassicality in such
cases with appropriate use of squeezing as a useful quantum resource,
which will be discussed elsewhere. In case of photon addition/subtraction
in DFS, a behavior analogous to that observed for higher-order antibunching
is observed, i.e., the depth of nonclassciality increases with the
photon addition while it decreases (increases) for small (large) values
of $\alpha$ (cf. Fig. \ref{HOSPS} (b) and (e)). Similar to the previous
case, nonclassicality can be observed to be present for larger values
of displacement parameter with photon addition/subtraction, while
increase in the value of Fock parameter has an opposite effect.

\subsection{Higher-order squeezing}

\begin{figure}
\begin{centering}
\begin{tabular}{ccc}
\includegraphics[scale=0.5]{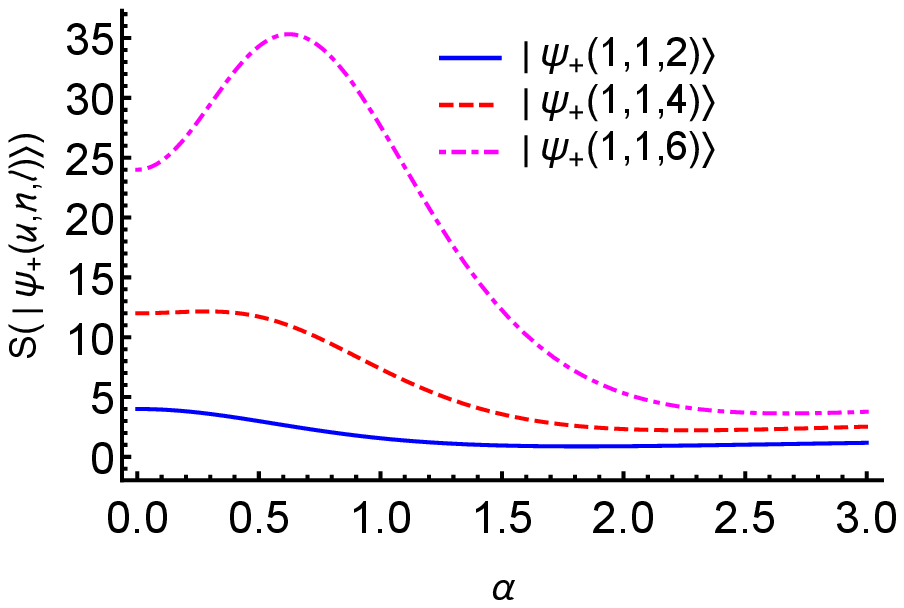}  & \includegraphics[scale=0.5]{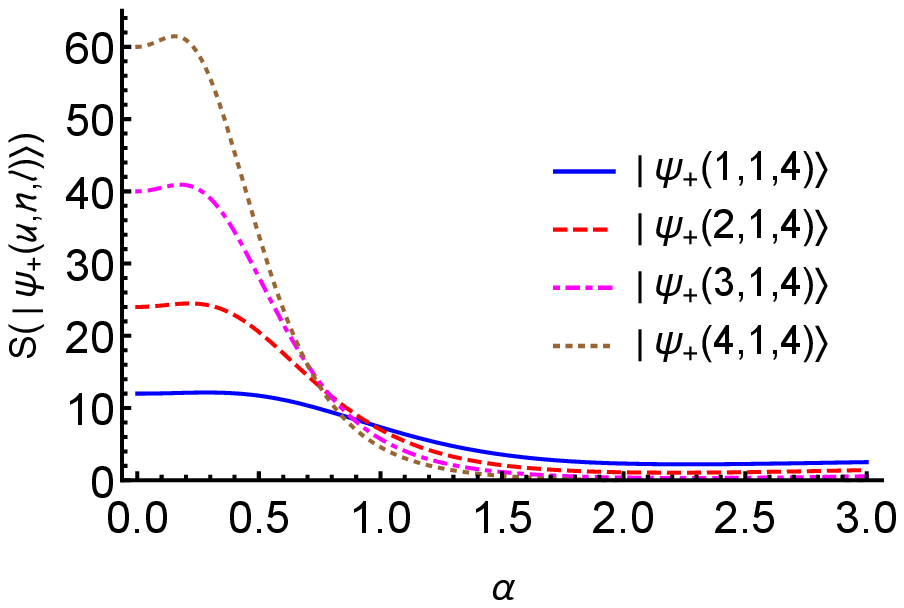}  & \includegraphics[scale=0.5]{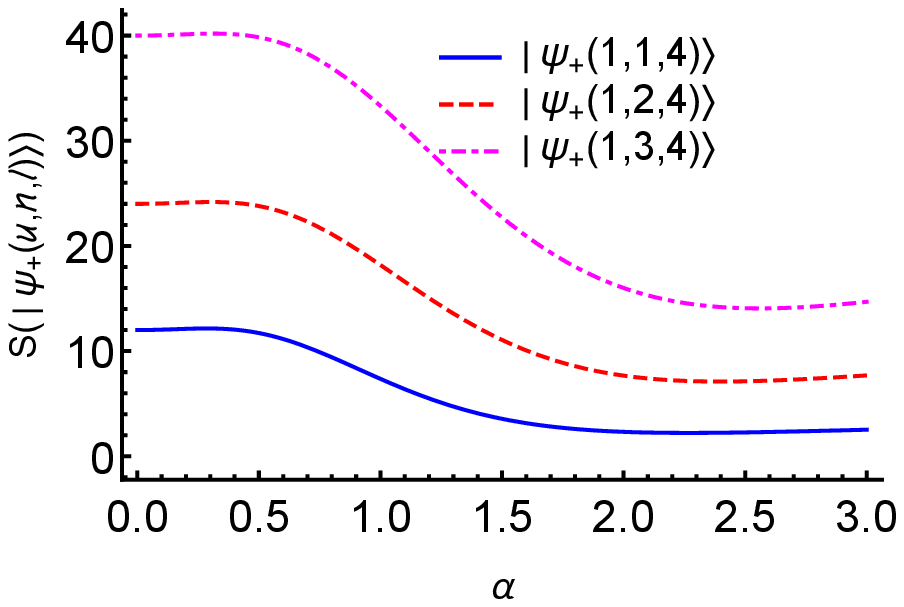} \tabularnewline
(a)  & (b)  & (c) \tabularnewline
\includegraphics[scale=0.5]{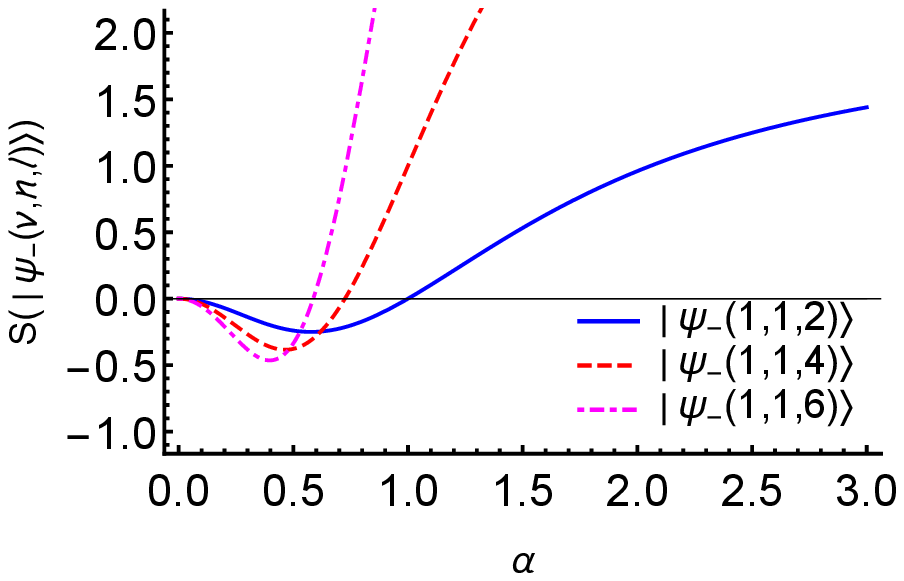}  & \includegraphics[scale=0.5]{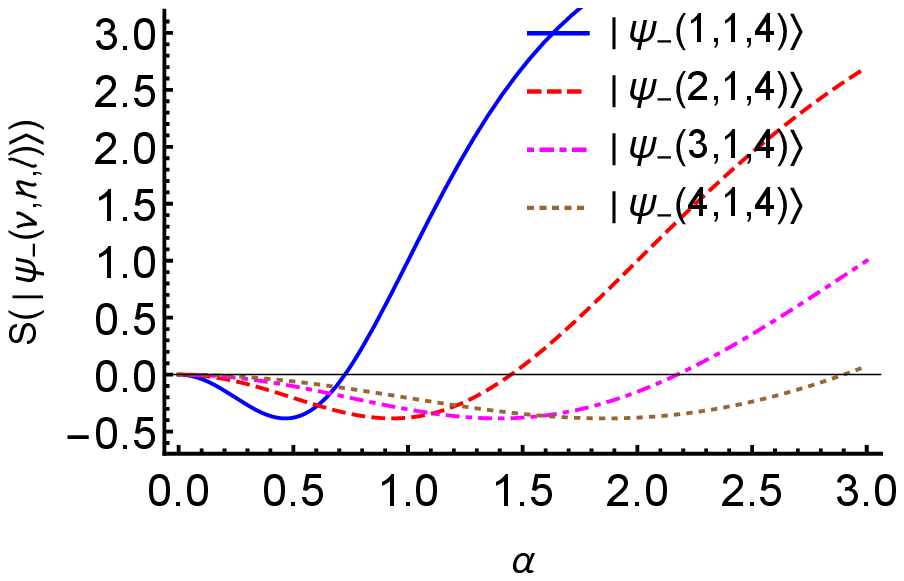}  & \includegraphics[scale=0.5]{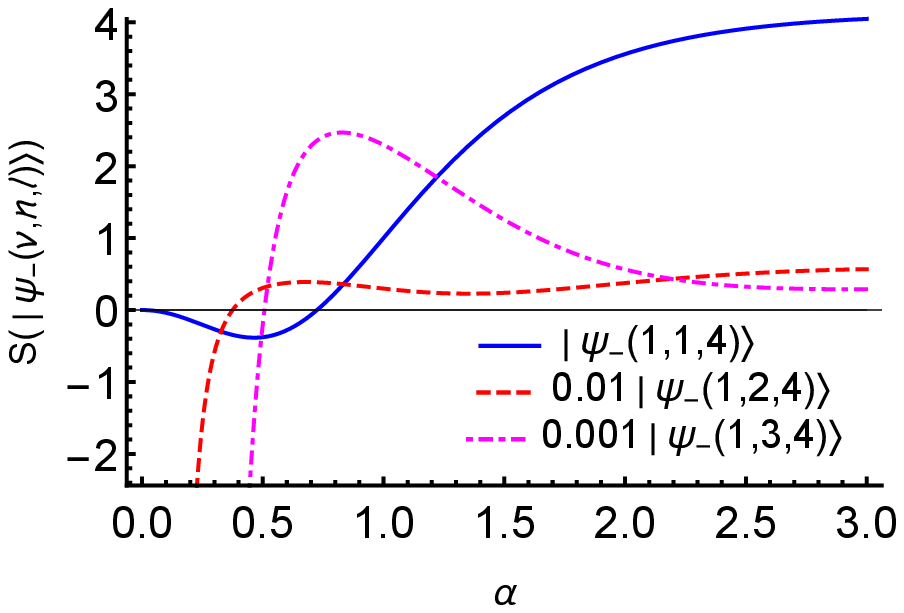} \tabularnewline
(d)  & (e)  & (f) \tabularnewline
\end{tabular}
\par\end{centering}
\caption{\label{fig:HOS} Illustration of the higher-order squeezing using
Hong-Mandel criterion as the function of displacement parameter. In
(a) and (d), dependence of the observed nonclassicality on different
orders ($l$) is shown for PADFS and PSDFS, respectively; while in
(b) and (e), the effect of variation in the number of photon added/subtracted
is shown in case of PADFS and PSDFS, respectively. In (c) and (f),
the variation due to change in the initial Fock state chosen to be
displaced is shown for PADFS and PSDFS, respectively. }
\end{figure}

\begin{figure}
\begin{centering}
\begin{tabular}{cc}
\includegraphics[scale=0.5]{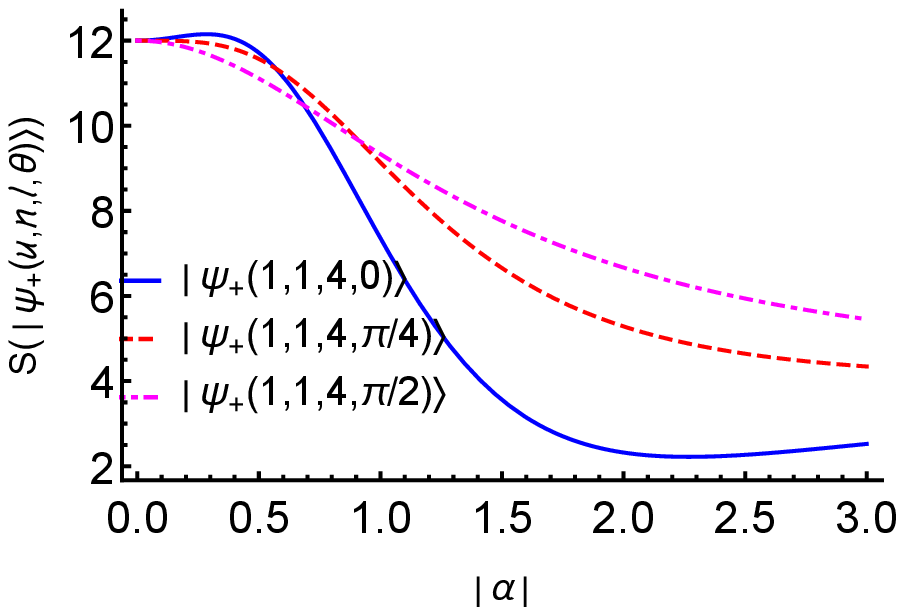}  & \includegraphics[scale=0.5]{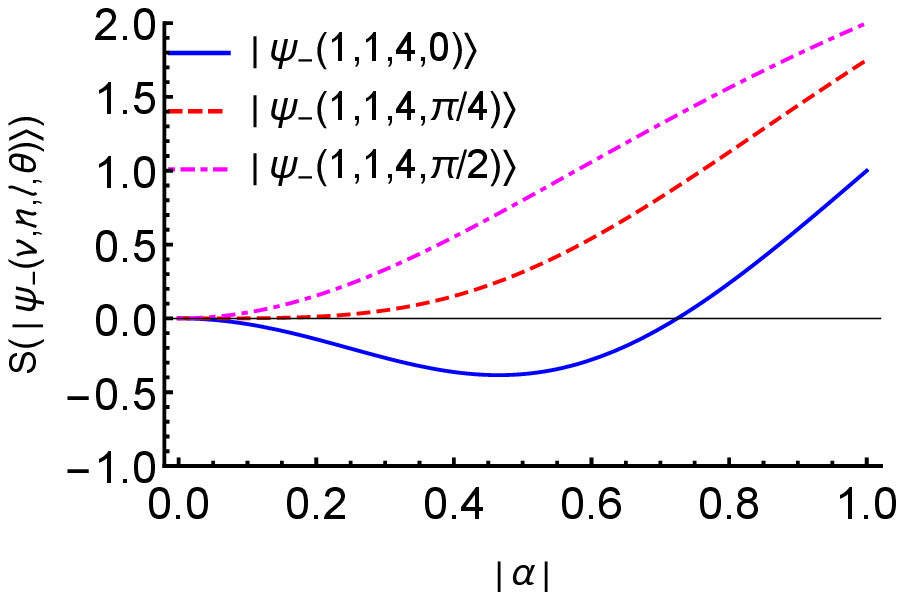} \tabularnewline
\end{tabular}
\par\end{centering}
\caption{\label{fig:HOS-diff-phase} Hong-Mandel type higher-order squeezing
for PADFS and PSDFS is shown dependent on the phase of the displacement
parameter $\alpha=|\alpha|\exp\left(i\theta\right)$ in (a) and (b),
respectively.}
\end{figure}

The squeezing of quadrature is defined in terms of variance in the
measured values of the quadrature (say, position or momentum) below
the corresponding value for the coherent state, i.e., minimum uncertainty
state. The higher-order counterpart of squeezing is studied in two
ways, namely Hong-Mandel and Hillery type squeezing \cite{hong1985higher,hong1985generation,hillery1987amplitude}.
Specifically, the idea of the higher-order squeezing originated from the pioneering
work of Hong and Mandel \cite{hong1985higher,hong1985generation},
who generalized the lower-order squeezing using the higher-order moment.
According to the Hong-Mandel criterion, the $l$th order squeezing
can be observed if the $l$th moment (for $l>2$) of a field quadrature
operator is less than the corresponding coherent state value. The
condition of Hong-Mandel type higher-order squeezing is given as follows
\cite{hong1985higher,hong1985generation}

\begin{equation}
S(l)=\frac{\langle(\Delta X)^{l}\rangle-\left(\frac{1}{2}\right)_{\frac{l}{2}}}{\left(\frac{1}{2}\right)_{\frac{l}{2}}}<0,\label{eq:Hong-Def}
\end{equation}
where $(x)_{l}$ is conventional Pochhammer symbol. The inequality
in Eq. (\ref{eq:Hong-Def}) can also be rewritten as 
\begin{equation}
\begin{array}{lcl}
\langle(\Delta X)^{l}\rangle & < & \left(\frac{1}{2}\right)_{\frac{l}{2}}=\frac{1}{2^{\frac{l}{2}}}(l-1)!!\end{array}\label{eq:Hong-def2}
\end{equation}
with 
\begin{equation}
\langle\left(\text{\ensuremath{\Delta}}\text{X}\right)^{l}\rangle=\sum\limits _{r=0}^{l}\sum\limits _{i=0}^{\frac{r}{2}}\sum\limits _{k=0}^{r-2i}\left(-1\right)^{r}\frac{1}{2^{\frac{l}{2}}}\left(2i-1\right)!^{2i}C_{k}{}^{l}C_{r}{}^{r}C_{2i}\langle\hat{a}^{\dagger}+\hat{a}\rangle^{l-r}\langle\hat{a}^{\dagger k}\hat{a}^{r-2i-k}\rangle,\label{eq:cond2.1}
\end{equation}
where $l$ is an even number, and the quadrature variable is defined
as $X=\frac{1}{\sqrt{2}}\left(a+a^{\dagger}\right)$. {{}
The analytical expressions of the nonclassicality witness of the Hong-Mandel
type higher-order squeezing criterion for PADFS and PSDFS can be obtained}
with the help of Eqs. (\ref{eq:PA-expepectation}), (\ref{eq:PS-expectation}),
and (\ref{eq:Hong-Def})-(\ref{eq:cond2.1}). We investigate the higher-order
squeezing and depict the result in Fig. \ref{fig:HOS}
assuming $\alpha$ to be real. Incidentally, we could not establish the
presence of higher-order squeezing phenomena in PADFS state Fig. \ref{fig:HOS}
(a)-(c). However, the depth of the higher-order squeezing witness
increases with order for small values of $\alpha$ as shown in Fig.
\ref{fig:HOS} (a), while for the higher values of the displacement
parameter, higher-order squeezing disappear much quicker (cf. Fig.
\ref{fig:HOS} (d)). With increase in the number of photons subtracted,
the presence of this nonclassicality feature can be maintained for
the higher values of displacement parameter as well (cf. Fig. \ref{fig:HOS}
(e)). In general, photon subtraction is a preferred mode for nonclassicality
enhancement as far as this nonclassciality feature is concerned. The
choice of the initial Fock state is also observed to be relevant as
the depth of squeezing parameter can be seen increasing with value
of the Fock parameter for PSDFS in the small displacement parameter
region (shown in Fig. \ref{fig:HOS} (f)), where this nonclassical
behavior is also shown to succumb to the higher values of Fock and
displacement parameters. Unlike the other nonclassicalities discussed
so far, the observed squeezing also depends on phase $\theta$ of
the displacement parameter $\alpha=|\alpha|\exp\left(i\theta\right)$
due to the second last term in Eq. (\ref{eq:cond2.1}). We failed
to observe this nonclassicality behavior in PADFS even by controlling
the value of the phase parameter (also shown in Fig. \ref{fig:HOS-diff-phase}
(a)). For PSDFS, the squeezing disappears for some particular values
of the phase parameter, while the observed squeezing is maximum for
$\theta=n\pi$ (see Fig. \ref{fig:HOS-diff-phase} (b)). It thus establishes
the phase parameter of the displacement operator as one more controlling
factor for nonclassicality in these engineered quantum states.

\subsection{$Q$ function}

\begin{figure}
\centering{}

\subfigure[]{\includegraphics[scale=0.5]{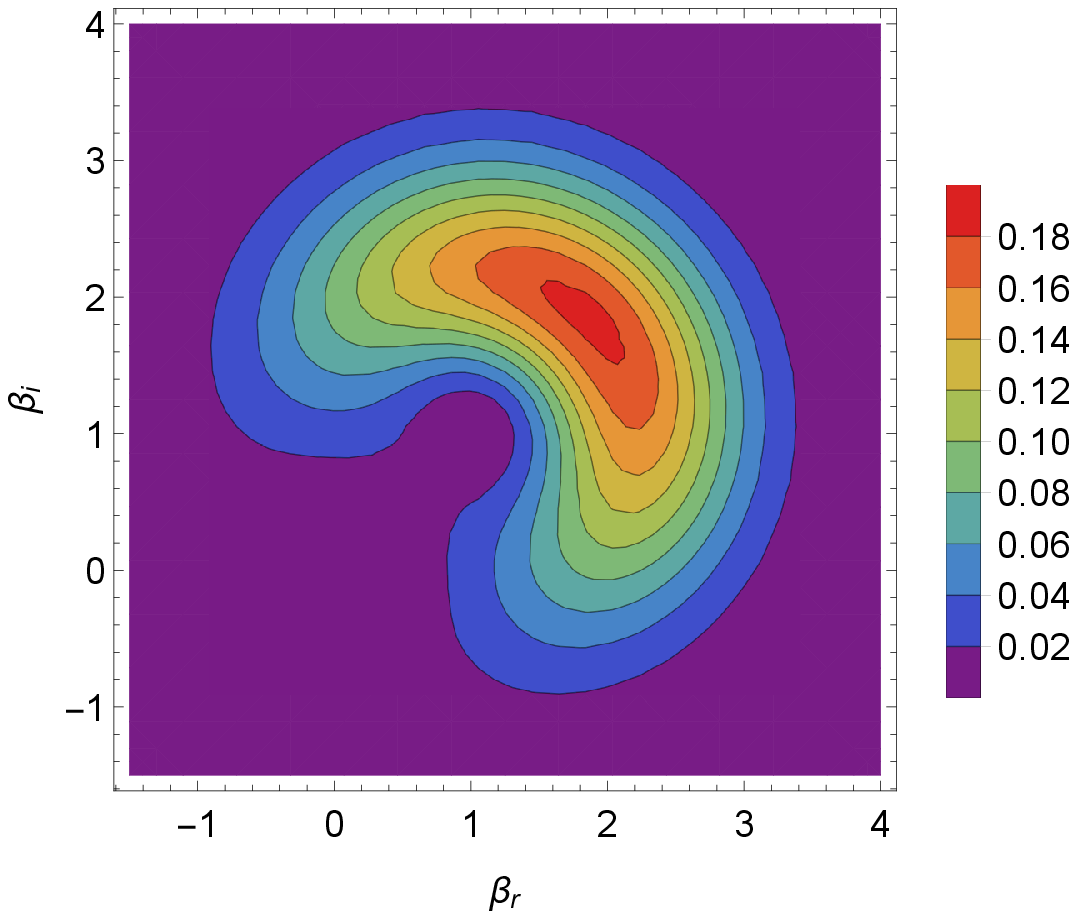}}
\quad{}\subfigure[]{\includegraphics[scale=0.5]{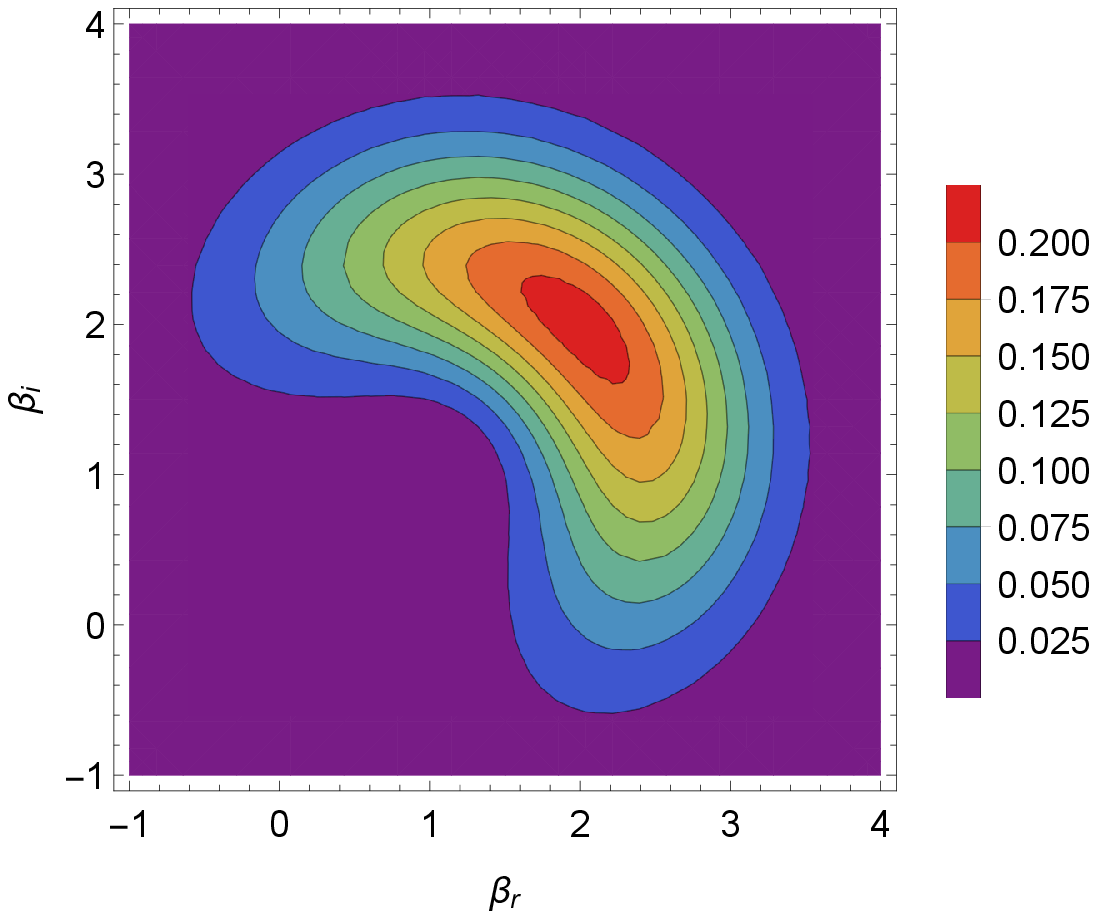}}\quad{}\subfigure[]{\includegraphics[scale=0.5]{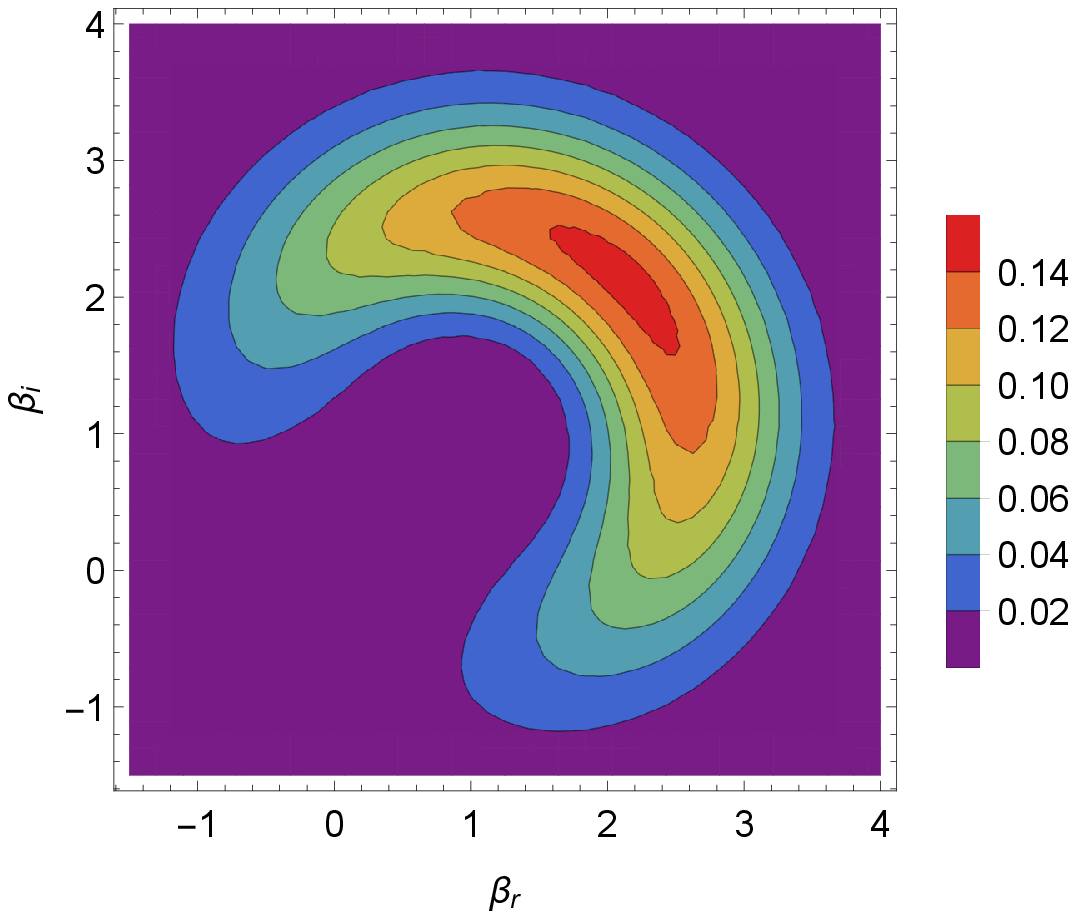}}\\
 \subfigure[]{\includegraphics[scale=0.5]{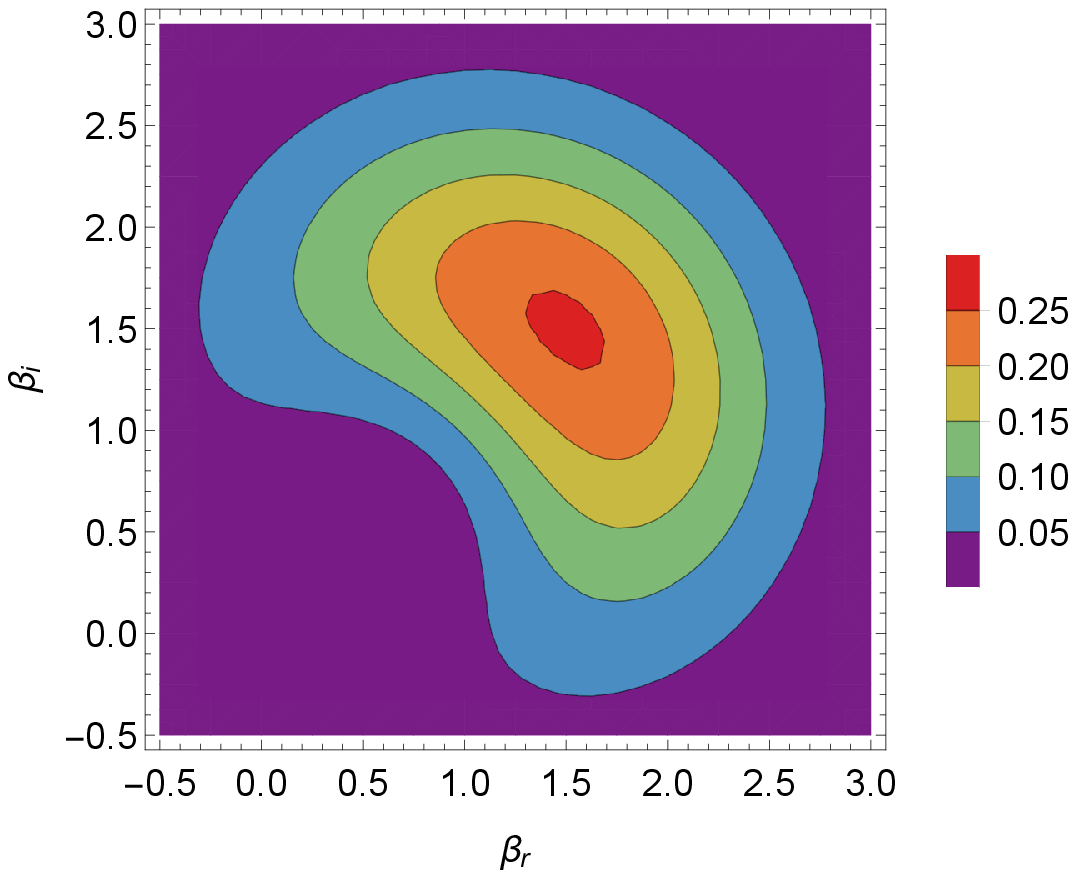}}\quad{}\subfigure[]{
\includegraphics[scale=0.5]{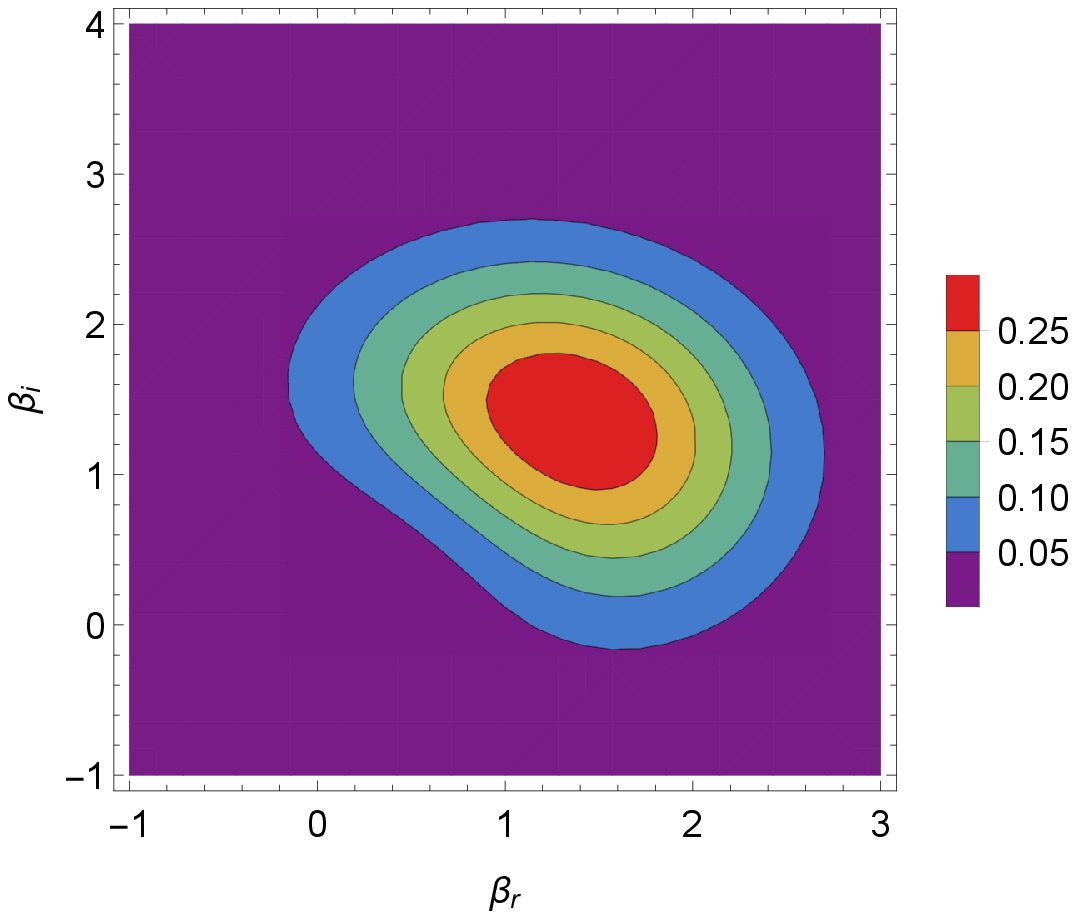}}\quad{}\subfigure[]{
\includegraphics[scale=0.5]{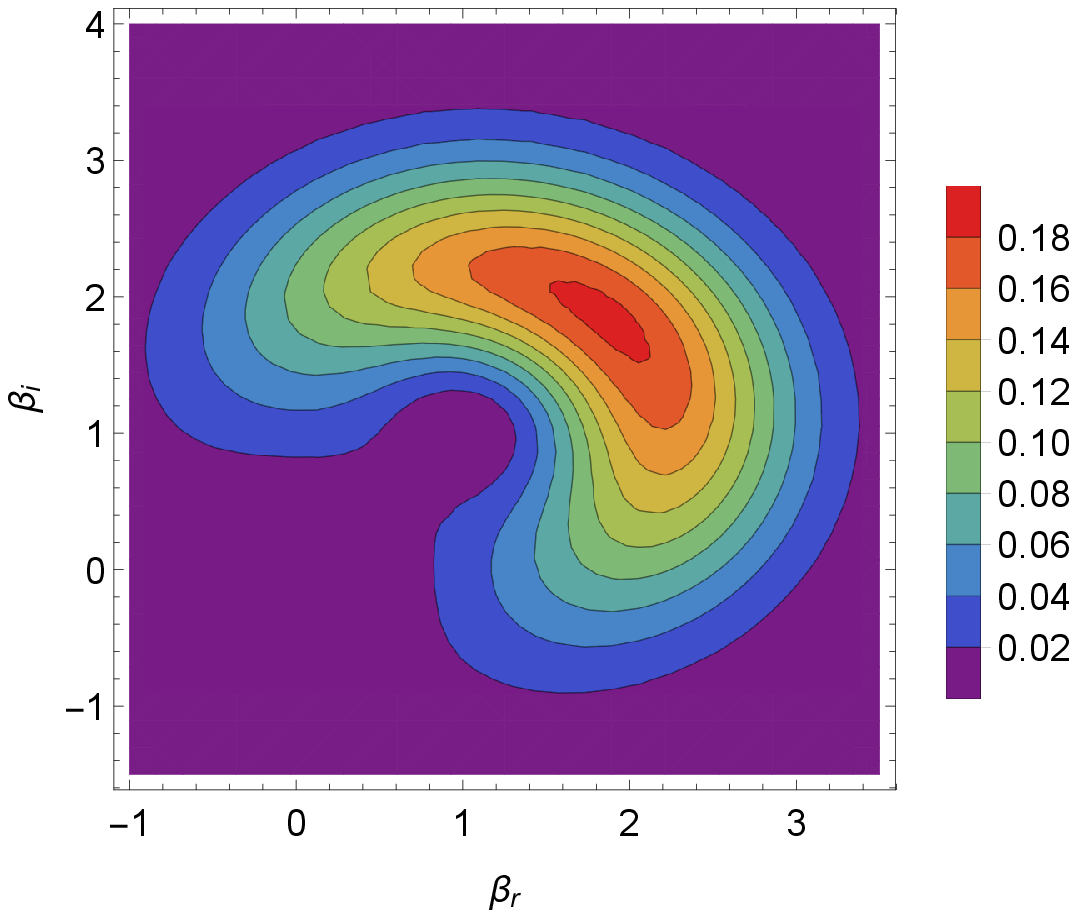}}\\

\caption{\label{fig:Q-function} (Color online) Contour plots of the $Q$ function
for (a) single photon added displaced Fock $|1\rangle$ state, (b)
two photon added displaced Fock $|1\rangle$ state, (c) single photon
added displaced Fock $|2\rangle$ state, (d) single photon subtracted
displaced Fock $|1\rangle$ state, (e) two photon subtracted subtracted
displaced Fock $|1\rangle$ state, (f) single photon subtracted displaced
Fock $|2\rangle$ state. In all cases, $\alpha=\sqrt{2}\exp\left(\frac{i\pi}{4}\right)$
is chosen. }
\end{figure}

Inability to give a phase space description of quantum mechanics is
exploited in terms of quasiprobability distributions (\cite{thapliyal2015quasiprobability}
and references therein) to use them as witnesses of nonclassicality.
These real and normalized quasiprobability distributions allow to
calculate the expectation value of an operator as any classical probability
distribution. One such quasiprobability distributions is $Q$ function
\cite{husimi1940some}, and zeros of this function are signature of
nonclassicality. $Q$ function \cite{husimi1940some} is defined as
\begin{equation}
Q=\dfrac{1}{\pi}\langle\beta|\rho|\beta\rangle,\label{eq:Q-function}
\end{equation}
where $|\beta\rangle$ is the coherent state. Using Eqs. (\ref{eq:PADFS})
and (\ref{eq:PSDFS}) in (\ref{eq:Q-function}), we obtain the analytic
expressions for the Husimi $Q$ function for PADFS and PSDFS as

\begin{equation}
\begin{array}{lcl}
Q_{+} & = & \frac{N_{+}^{2}}{\pi}\frac{\exp\left[-\mid\beta\mid^{2}\right]}{n!}\sum\limits _{p,p'=0}^{n}{n \choose p}{n \choose p'}(-\alpha^{\star})^{(n-p)}(-\alpha)^{(n-p')}\exp\left[-\mid\alpha\mid^{2}\right]\sum\limits _{m,m^{\prime}=0}^{\infty}\frac{\alpha^{m}(\alpha^{\star})^{m^{\prime}}\beta^{(m^{\prime}+p'+u)}(\beta^{\star})^{(m+p+u)}}{m!m^{\prime}!}\end{array}\label{eq:Q-PADFS}
\end{equation}
and 
\begin{equation}
\begin{array}{lcl}
Q_{-} & = & \frac{N_{-}^{2}}{\pi}\frac{\exp\left[-\mid\beta\mid^{2}\right]}{n!}\sum\limits _{p,p'=0}^{n}{n \choose p}{n \choose p'}(-\alpha^{\star})^{(n-p)}(-\alpha)^{(n-p')}\exp\left[-\mid\alpha\mid^{2}\right]\\
 & \times & \sum\limits _{m,m^{\prime}=0}^{\infty}\frac{\alpha^{m}(\alpha^{\star})^{m^{\prime}}\beta^{(m^{\prime}+p'-v)}(\beta^{\star})^{(m+p-v)}(m+p)!(m^{\prime}+p')!}{m!m^{\prime}!(m+p-v)!(m^{\prime}+p'-v)!},
\end{array}\label{eq:Q-PSDFS}
\end{equation}
respectively. We failed to observe the nonclassical features reflected
beyond moments based nonclassicality criteria through a quasiprobability
distribution, i.e., zeros of the $Q$ function. We have shown the
$Q$ function in Fig. \ref{fig:Q-function}, where the effect of photon
addition/subtraction and the value of Fock parameter on the phase
space distribution is shown. Specifically, it is observed that the
value of Fock parameter affects the quasidistribution function more
compared to the photon addition/subtraction.

\subsection{Agarwal-Tara criterion}

We will further discuss one more moments based criteria of nonclassicality.
This particular criterion was introduced by Agarwal and Tara \cite{agarwal1992nonclassical},
and it can be written in a matrix form. The criterion is defined as
\begin{equation}
A_{3}=\dfrac{\det m^{(3)}}{\det\mu^{(3)}-\det m^{(3)}}<0,\label{eq:Agarwal}
\end{equation}
where 
\[
m^{(3)}=\begin{bmatrix}1 & m_{1} & m_{2}\\
m_{1} & m_{2} & m_{3}\\
m_{2} & m_{3} & m_{4}
\end{bmatrix}
\]
and 
\[
\mu^{(3)}=\begin{bmatrix}1 & \mu_{1} & \mu_{2}\\
\mu_{1} & \mu_{2} & \mu_{3}\\
\mu_{2} & \mu_{3} & \mu_{4}
\end{bmatrix}.
\]
The matrix elements are defined as $m_{i}=\langle\hat{a}^{\dagger i}\hat{a}^{i}\rangle$
and $\mu_{j}=(\langle\hat{a}^{\dagger}\hat{a}\rangle)^{j}=(m_{1})^{j}$.

\begin{figure}
\centering{}

\subfigure[]{\includegraphics[scale=0.6]{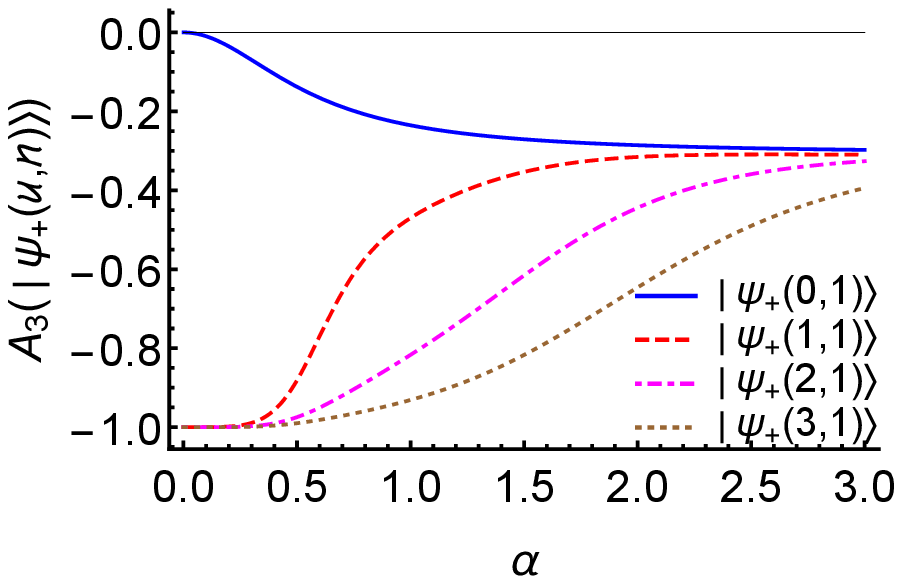}}
\quad{}\quad{}\subfigure[]{ \includegraphics[scale=0.6]{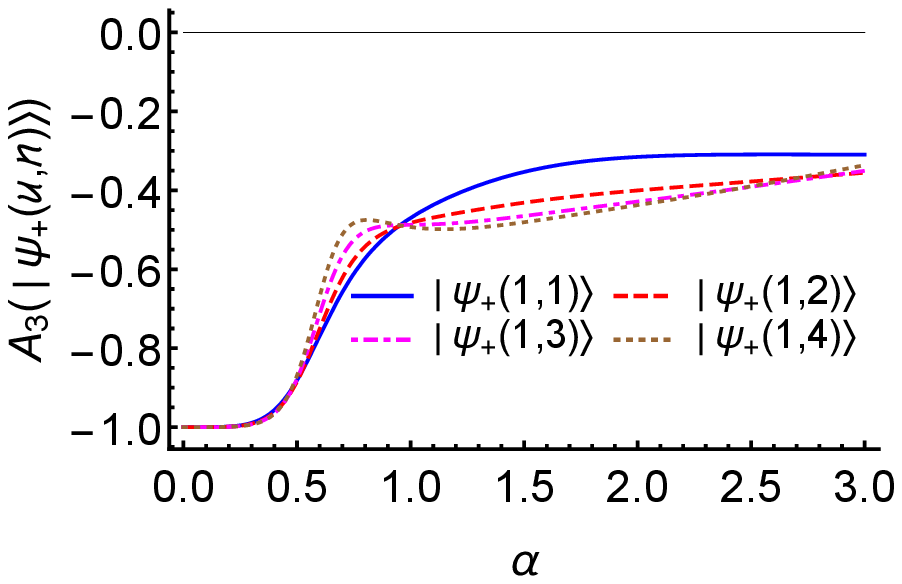}}\\
 \subfigure[]{\includegraphics[scale=0.6]{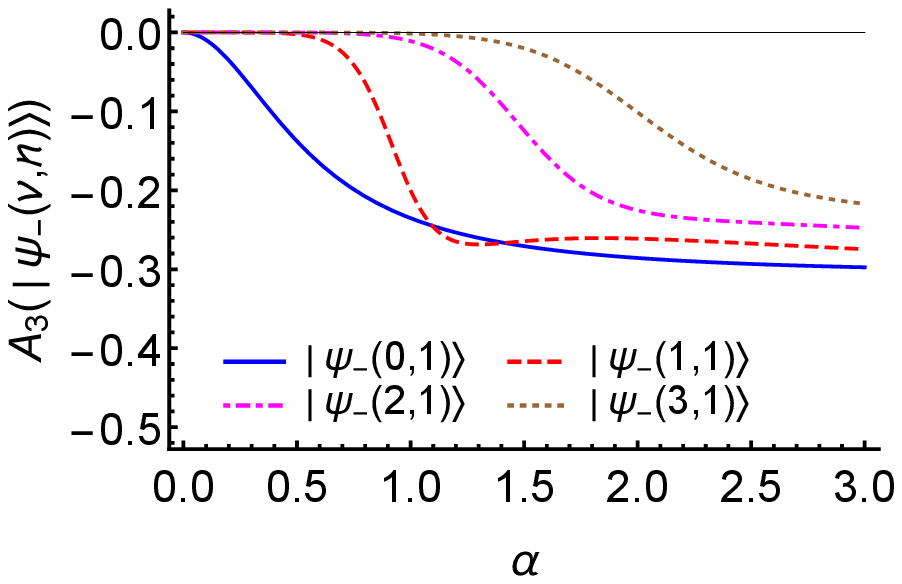}}\quad{}\quad{}\subfigure[]{
\includegraphics[scale=0.6]{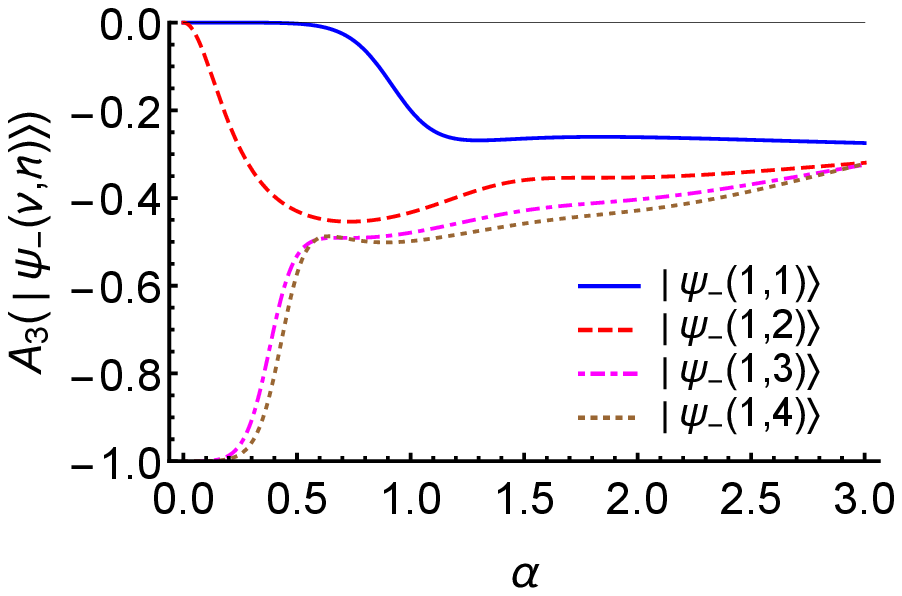}}\\

\caption{\label{fig:A3}(Color online) Variation of Agarwal-Tara parameter
with $\alpha$ for PADFS and PSDFS is shown in (a)-(b) and (c)-(d),
respectively. Specifically, the effect of photon addition/subtraction
(in (a) and (c)) and the choice of Fock state ((b) and (d)) on the
presence of nonclassicality in PADFS and PSDFS is illustrated.}
\end{figure}

The analytic expression of $A_{3}$ parameter defined in Eq. (\ref{eq:Agarwal})
{can} be obtained for PADFS and PSDFS using Eqs. (\ref{eq:PA-expepectation})
and (\ref{eq:PS-expectation}). The nonclassical properties of the
PADFS and PSDFS using Agarwal-Tara criterion are investigated, and
the corresponding results are depicted in Fig. \ref{fig:A3}, which
shows highly nonclassical behavior of the {states
generated by engineering}. Specifically, the negative part of the
curves, which is bounded by -1, ensures the existence of the nonclassicality.
From Fig. \ref{fig:A3}, it is clear that $A_{3}$ is 0 (-1) for the
displacement parameter $\alpha=0$ because then DFS, PADFS, and PSDFS
reduce to Fock state, and $A_{3}=0$ (-1) for the Fock state parameter
$n=0,\,1$ $\left(n>1\right)$. Nonclassicality reflected through
$A_{3}$ parameter increases (decreases) with photon addition (subtraction)
(shown in Fig. \ref{fig:A3} (a) and (c)). In contrast, the Fock parameter
has a completely opposite effect that it leads to decrease (increase)
in the observed nonclassicality for PADFS (PSDFS), which can be seen
in Fig. \ref{fig:A3} (b) and (d). However, for larger values of displacement
parameter, the depth of nonclassicality illustrated through this parameter
can again be seen to increase (cf. Fig. \ref{fig:A3} (b)).

\subsection{Klyshko's Criterion}

Klyshko introduced a criterion \cite{klyshko1996observable} to investigate
the nonclassical property using only three successive photon-number {{}
probabilities}. In terms of the photon-number probability $p_{m}=\langle m|\rho|m\rangle$
of the state with density matrix $\rho$, the Klyshko's criterion
in the form of an inequality can be written as 
\begin{equation}
B(m)=(m+2)p_{m}p_{m+2}-(m+1)\left(p_{m+1}\right)^{2}<0.\label{eq:Klyshko}
\end{equation}
The analytic expression for the $m$th photon-number distribution
for PADFS and PSDFS can be calculated (using $q=r=1$) from Eqs. (\ref{eq:PA-expepectation})
and (\ref{eq:PS-expectation}), respectively.

\begin{figure}
\centering

\subfigure[]{\includegraphics[scale=0.6]{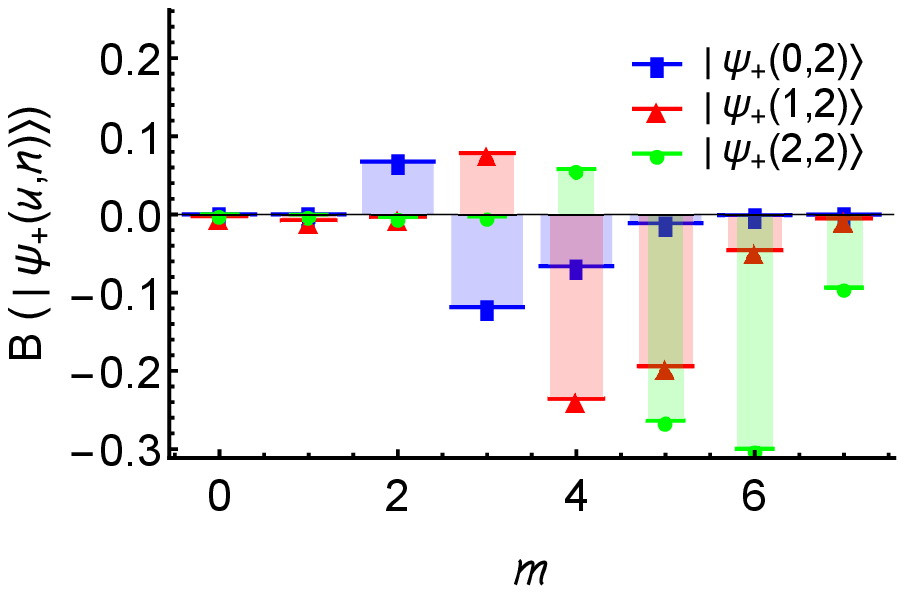}}\quad{}\quad{}\subfigure[]{
\includegraphics[scale=0.6]{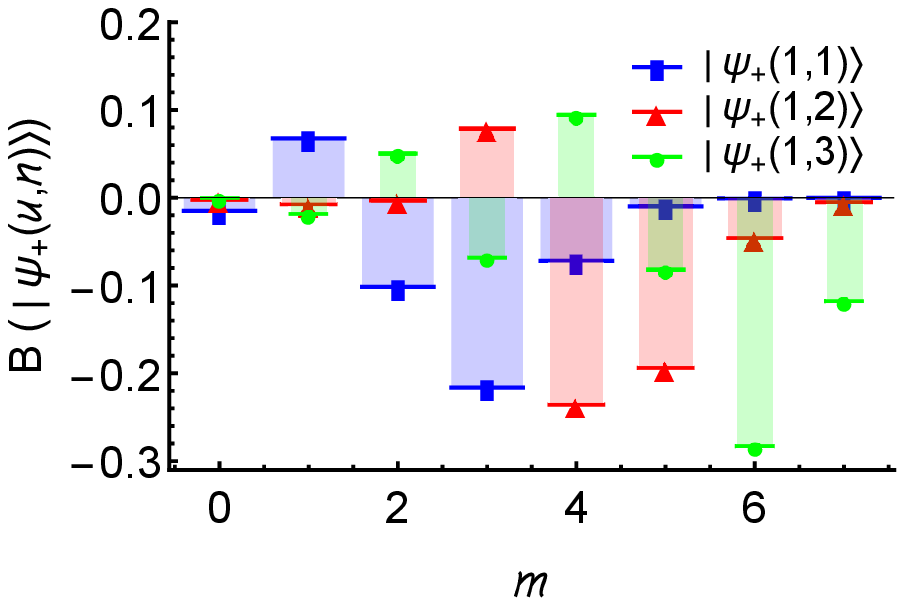}}\\
 \subfigure[]{\includegraphics[scale=0.6]{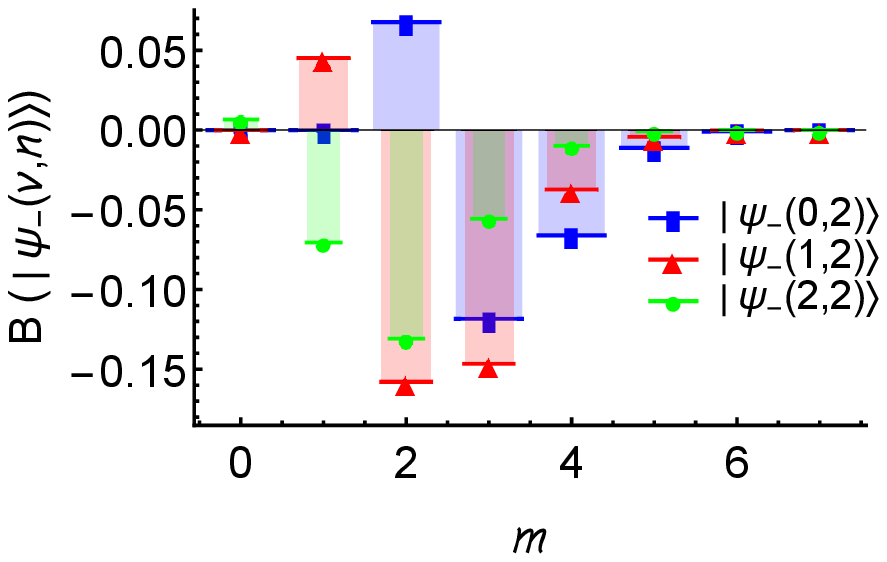}}\quad{}\quad{}\subfigure[]{
\includegraphics[scale=0.6]{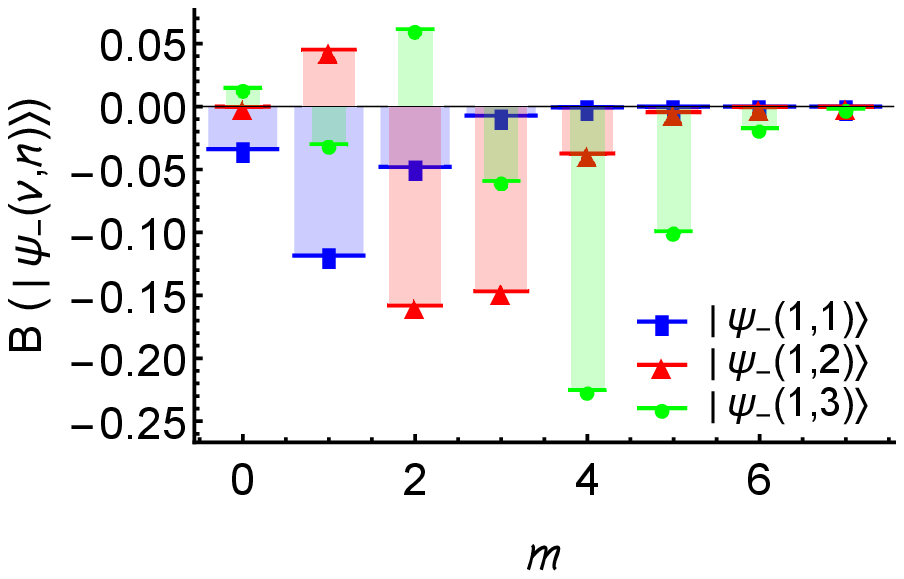}}\\

\caption{\label{fig:Klyshko} (Color online) Illustration of the Klyshko's
criterion. Variation of $B(m)$ with respect to $m$ (a) and (c) for
different values of the number of the photon additon/subtraction for
PADFS and PSDFS, respectively; (b) and (d) for different values of
the number of the Fock state parameter for PADFS and PSDFS, respectively.
Here, we have chosen $\alpha=1$ in all cases. }
\end{figure}

The advantage of the Klyshko's criterion over any other existing moments
based criteria is that a very small amount of information is required.
Specifically, probability of only three successive photon numbers
is sufficient to investigate the nonclassical property. The negative
values of $B(m)$ serve as the witness of nonclassicality. Klyshko's
criterion in Eq. (\ref{eq:Klyshko}) is derived analytically and the
corresponding nonclassical properties for both PADFS and PSDFS are
investigated (cf. Fig. \ref{fig:Klyshko}). Specifically, the negative
values of $B(m)$ are observed for different values of $m$ in case
of photon addition and subtraction (cf. Fig. \ref{fig:Klyshko} (a)
and (c)) being the signature of nonclassicality induced via independent
operations. Additionally, one can also visualize that due to the photon
addition (subtraction) the negative peaks in the values of $B(m)$
shift to higher (lower) photon number regime. A similar observation
is obtained for different values of the Fock state parameter for the
PADFS and PSDFS, where the negative values of the witness of nonclassicality
get amplified towards higher photon number regime and becomes more
negative, and the corresponding results are shown in Fig. \ref{fig:Klyshko}
(b) and (d), respectively. This further establishes the relevance
of operations, like photon addition, subtraction, and starting with
Fock states in inducing nonclassicality in the engineered quantum
states.

\section{Conclusions \label{sec:Conclusions}}

The only Fock state that does not show any nonclassical feature is
the vacuum state \cite{miranowicz2015statistical}, and displacement
operator preserves its classicality. All the rest of the Fock states
are maximally nonclassical and they are shown to remain nonclassical
even after application of a displacement operator. Here, we set ourselves
a task: What happens when the displacement operator applied on a Fock
state is followed by addition or subtraction of photon(s)? Independently,
photon addition and subtraction are already established as nonclassicality
inducing operations in case of displaced vacuum state (i.e., coherent
state). We have established that photon addition/subtraction is not
only nonclassicality inducing operation, it can also enhance the nonclassicality
present in the DFS. It's expected that these operations would increase
the amount of nonclassicality (more precisely, the depth of the nonclassicality
witnessing parameter) present in other nonclassical states, too. There
is one more advantage of studying the nonclassical features of PADFS
and PSDFS. These states can be reduced to a class of quantum states,
most of which are already experimentally realized and found useful
in several applications. Inspired from the available experimental
results, we have also proposed optical designs for generation of PADFS
and PSDFS from squeezed vacuum state.

To analyze the nonclassical features of the engineered final states,
i.e., PADFS and PSDFS, we have used a set of moments based criteria,
namely Mandel $Q_{M}$ parameter, Agarwal-Tara $A_{3}$ parameter,
criteria for higher-order antibunching, sub-Poissonian photon statistics,
and squeezing. In addition, the nonclassical features have been investigated
through Klyshko's criterion and a quasiprobability distribution\textendash $Q$
function. The states of interest are found to show a large variety
of nonclassical features as all the nonclassicality witnesses used
here (except $Q$ function) are found to detect the nonclassicality.

This study has revealed that the amount of nonclassicality in PADFS
and PSDFS states can be controlled by the Fock state parameter, displacement
parameter, the number of photons added or subtracted. In general,
the amount of nonclassicality with respect to the witness used here
is found to increase with the number of photons added/subtracted,
while smaller values of Fock state and displacement parameters are
observed to be preferable for the presence of various nonclassical
features. On some occasions, nonclassicality has also been observed
to increase with the Fock state parameter, while larger values of
displacement parameter always affect the nonclassicality adversely.
Most of the nonclassicality criteria used here, being moments-based
criteria, could not demonstrate the effect of phase parameter of the
displacement parameter. Here, higher-order squeezing witness and $Q$
function are found dependent on the phase of the displacement parameter.
However, only higher-order squeezing criterion was able to detect
nonclassicality, and thus established that this phase parameter can
also be used to control the amount of nonclassicality.

Further, in the past, it has been established that higher-order nonclassicality
criteria have an advantage in detecting weaker nonclassicality. We
have also shown that the depth of nonclassicality witness increases
with order of nonclassicality thus providing an advantage in the experimental
characterization of the observed nonclassical behavior. Finally, we
feel that this study can be extended to investigate the non-Gaussianity
of the studied state and hope that the observed nonclassicality in
these states would find applications in quantum information processing
tasks.

\textbf{Acknowledgment:} A.P. and N.A. thank the Department of Science
and Technology (DST), India, for support provided through the DST
project No. EMR/2015/000393. S.B and V. N. thank CSIR, New Delhi for
support through the project No. 03(1369)/16/EMR-II.

\bibliographystyle{elsarticle-num}
\bibliography{biblio,priya}

\end{document}